 \documentclass[preprint2]{aastex}

\slugcomment{Submitted to AJ (revised version)} 
 
\shorttitle{Criteria to classify T Tauri stars and BDs} 
\shortauthors{Barrado y Navascu\'es \& Mart\'{\i}n} 
 
\begin{document} 
 
\title{An empirical criterion to classify T Tauri stars and substellar  
analogs using low-resolution optical spectroscopy} 
 
\author{David Barrado y Navascu\'es} 
\affil{Laboratorio de Astrof\'{\i}sica Espacial y F\'{\i}sica Fundamental, 
           INTA,  P.O. Box 50727, E-28080 Madrid, Spain} 
 
              \email{barrado@laeff.esa.es} 
 
\and 
 
\author{Eduardo L. Mart\'{\i}n\footnote{also at 
Instituto de Astrof\'{\i}sica de Canarias, 38200 La Laguna, Spain}} 
\affil{Institute of Astronomy. University of Hawaii at Manoa.  
             2680 Woodlawn Drive, Honolulu, HI 96822, USA.}

\begin{abstract} 
We have compiled and studied photometric and spectroscopic 
data published in the literature of several 
star forming regions and young open clusters 
(Orion, Taurus, IC348, Sco-Cen Complex, Chamaeleon~I, TW Hya association, 
$\sigma$ Orionis cluster, IC2391, $\alpha$ Per cluster  and the Pleiades). 
Our goal was to seek the definition of a simple empirical criterion to 
classify  stars or  brown dwarfs which are accreting matter from a disk on 
the sole basis of low-resolution optical spectroscopic data. 
We show that using H$\alpha$ equivalent widths and spectral types we can 
statistically classify very young stars and brown dwarfs as classical 
T Tauri stars and substellar analogs. As a boundary between accreting and 
non accreting objects, we use the saturation limit of chromospheric activity 
 at  Log \{L(H$\alpha$)/L(bol)\}=$-$3.3 (determined in the open clusters). 
We discuss the uncertainties in the classification scheme due 
to the occurrence of flares. We have used this 
spectroscopic empirical criterion to classify objects found in the literature, 
and we compute the fraction of accreting objects in several 
star forming regions. The fraction of accreting objects appears to decrease 
from about 50\% to about 5\% from 1~Myr to 10~Myr for both stars and brown dwarfs. 
\end{abstract} 


%%%%%%%%%%%%%%%%%%%%%%%%%%%%%%%%%%%%%%%%%%%%%%%%%%%%%%%%%%%%%%%%%%%% 
%%%%%%%%%%%%%%%%%%%%%%%%%%%%%%%%%%%%%%%%%%%%%%%%%%%%%%%%%%%%%%%%%%%% 
%%%%%%%%%%%%%%%%%%%%%%%%%%%%%%%%%%%%%%%%%%%%%%%%%%%%%%%%%%%%%%%%%%%% 
%%%%%%%%%%%%%%%%%%%%%%%%%%%%%%%%%%%%%%%%%%%%%%%%%%%%%%%%%%%%%%%%%%%% 
%%%%%%%%%%%%%%%%%%%%%%%%%%%%%%%%%%%%%%%%%%%%%%%%%%%%%%%%%%%%%%%%%%%% 
% 
%                      TEXT 
% 
%%%%%%%%%%%%%%%%%%%%%%%%%%%%%%%%%%%%%%%%%%%%%%%%%%%%%%%%%%%%%%%%%%%% 
%%%%%%%%%%%%%%%%%%%%%%%%%%%%%%%%%%%%%%%%%%%%%%%%%%%%%%%%%%%%%%%%%%%% 
%%%%%%%%%%%%%%%%%%%%%%%%%%%%%%%%%%%%%%%%%%%%%%%%%%%%%%%%%%%%%%%%%%%% 
%%%%%%%%%%%%%%%%%%%%%%%%%%%%%%%%%%%%%%%%%%%%%%%%%%%%%%%%%%%%%%%%%%%% 
%%%%%%%%%%%%%%%%%%%%%%%%%%%%%%%%%%%%%%%%%%%%%%%%%%%%%%%%%%%%%%%%%%%% 

%%%%%%%%%%%%%%%%%%%%%%%%%%%%%%%%%%%%%%%%%%%%%%%%%% 
%%%%%%%%%%%%%                   %%%%%%%%%%%%%%%%%% 
%%%%%%%%%%%%%    Section        %%%%%%%%%%%%%%%%%% 
%%%%%%%%%%%%%                   %%%%%%%%%%%%%%%%%% 
%%%%%%%%%%%%%%%%%%%%%%%%%%%%%%%%%%%%%%%%%%%%%%%%%% 

\section{Introduction}

A classical T~Tauri star (CTTS) is a pre-main sequence (PMS) star 
which  presents a characteristic phenomenology which, in general,  
includes the following: 
i) An emission line spectrum, including, in the visible,
  lines such as  
the Balmer series with very intense  H$\alpha$; He{\sc i} D$_1$,D$_2$,D$_3$ and  
He{\sc i} 6678 \AA; 
Ca{\sc ii} H \& K and Ca{\sc ii} infrared triplet at 8498, 8542 and 8662 \AA; etcetera. 
These emission lines (in particular H$\alpha$) are quite broad (several  
hundred km/s at FWHM), and  can present asymmetries. 
ii) The presence of forbidden --generally blue-shifted  narrow-- lines 
(for example, [O{\sc i}]6300\&6364, [N{\sc ii}]6458\&6581, [S{\sc ii}]6717\&6731). 
iii) Photospheric continuum excesses, specially in the ultraviolet,  
the blue part of the optical range and the infrared (Class II spectral energy distribution).
Not all of them appear simultaneously. 

There are different proposed criteria to define CTTS,
and they do not always agree with each other. 
The simplest criterion is based on the equivalent width 
of H$\alpha$ (W(H$\alpha$)). Critical values between 5 and 20 \AA{ } have been put forward. 
Mart\'{\i}n (1998)
 suggested  a dependence with the spectral type: 
 5 \AA{ } for spectral types earlier 
than M0, 10 \AA{ } for M0--M2,  and 20 \AA{ } for later 
spectral types.  

The canonical interpretation of  CTTS phenomenology 
 states that they possess a circumstellar disk, 
containing a significant fraction of the mass of the star, which is the source 
of the infrared excess,   and a strong magnetosphere 
(see Appenzeller \& Mundt 1989 for a review). 
Material from the disk would be magnetically accreted onto the star, 
 in a channeled flux, 
creating a hot spot, the source of the continuum excesses in the 
 blue side of the spectrum. 
The  broad, asymmetric H$\alpha$ (or other  hydrogen lines) 
 is produced 
by this material falling from the disk.  
In general, for permitted lines,  blue-shifted absorptions  
denote strong winds, whereas red-shifted absorptions indicate infalls. 
Finally, the forbidden lines  
are produced in outflows, which are --as this interpretation goes-- 
 jets perpendicular to the disk, and are  driven by the accretion. 
For more details, see Shu et al$.$ (1994),  
Hartmann et al$.$ (1994) and Hartmann  (1998).

Weak-line T~Tauri stars (WTTS) lack most of 
the observational features of CTTS. 
They show H$\alpha$ in emission 
but  with equivalent widths  smaller than in
 the case of CTTS (Herbig \& Bell 1988). 
Some WTTS show infrared excess, but not
 ultraviolet excess or optical veiling.  
Naked T Tauri stars (NTTS) are equivalent to 
the WTTS but without infrared excess  (Walter 1986).
It is assumed that WTTS are in a more advanced 
stage of their evolution, with 
no circumstellar accretion disks. 
Hence, the origin of the emission in H$\alpha$ 
should be purely chromospheric.
WTTS are still very young ($<$10 Myr), 
and their interior has not reached a 
temperature hot enough to destroy lithium
 which burns at about 2.5$\times$10$^6$ K. 
Finally, post-T~Tauri stars (PTTS) represent
 the subsequent PMS evolution of the 
previous two phases. Together with rapid
 rotation and high activity levels, 
they have already started the lithium depletion
 (Mart\'\i n 1997, 1998).

Brown dwarfs (BD) are substellar objects which are unable to 
 settle on the main sequence because their interior reaches a degenerate state 
(Hayashi \& Nakano 1963; Kumar 1963). For solar metallicity, current models 
predict that the borderline between stars and BDs is at 0.072 M$_\odot$ 
(Burrows et al$.$ 1997; Baraffe et al$.$ 1998; Chabrier et al$.$ 2000).  
BDs are low-mass analogs to PMS stars  
because they are fully convective, gravitationally contracting objects.  
There is  a continuity from the stellar to the 
 substellar regime in a color-magnitude diagram, a continuous dwarf sequence. 
In fact, very low mass (VLM) stars need tens of billion of years to reach the MS 
 (longer than the age of the Universe).  
Observationally, the substellar limit is located at spectral type of M6 in 
the Pleiades cluster (Mart\'{\i}n et al. 1996).
 This spectral type boundary between stars and BDs 
is also used in the literature for very young ages 
(Luhman et al$.$ 1998).  

Since the discovery of the first confirmed BDs in the Pleiades cluster     
(Rebolo et al$.$ 1995; Basri et al$.$ 1996), 
an avalanche of discoveries  and theoretical work  has taken place. 
The standard formation mechanism of BDs is that they come from small 
 cores produced by the fragmentation of molecular clouds 
--star formation does not know about the H-burning substellar limit.  
Recent theoretical work indicates that very small fragments can be 
 produced by turbulence in molecular clouds  
(Padoan \& Nordlund 2002). If BDs are formed in a similar manner as stars, 
 it seems natural to expect that some very young BDs could share 
 the same properties as CTTS. In particular, they could have  
circum-substellar disks with active accretion leading to an emission 
 line spectrum and continuum excesses.  
In fact, during the last 2 years some initial  
reports indicate that a few of them,  indeed, 
have disks (Mart\'{\i}n et al$.$ 2001a; Natta \& Testi 2001; Natta et al$.$ 2002;  
Testi et al$.$ 2002; Jayawardhana et al$.$ 2002ab, 2003ab; 
Barrado y Navascu\'es et al$.$ 2002, 2003; White \& Basri 2003). 

The main aim of this paper is to explore an empirical criterion based on low-resolution 
optical spectroscopy, which 
allows to identify accretion phenomena by distinguishing 
between CTTS, WTTS and their substellar analogs. We adopt the  
following definitions: substellar classical T Tauri analog (SCTTA),  
which is a substellar-mass object that shows emission lines  
indicative of active accretion similar to those observed in CTTSs; 
substellar weak-line T Tauri analog (SWTTA), which shows 
 chromospheric line emission similar to that of WTTS.  
We find that SCTTAs and SWTTAs can be classified statistically from  
low-resolution optical spectroscopy using spectral types and 
W(H$\alpha$). 
Another criterion, proposed by White \& Basri (2003), is based on the width of this line, 
but its application demands high resolution spectroscopy which is more difficult to 
obtain, particularly for intrinsically faint substellar objects. 
Our study extends the classical/weak-line 
T~Tauri classification into the substellar regime. 
We have applied our criterion to derive the ratio of SCTTA/SWTTA 
 in different star-forming regions. This ratio allows  
us to discuss the dependence of accretion on age, mass and location. 

In section 2, we discuss some of the properties of the
 CTTS and WTTS population of very young clusters and star
 forming regions. It also presents an analysis 
of the H$\alpha$ equivalent width as a the main empirical  
criterion to classify CTTS, WTTS, SCTTA and SWTTA.  
In Section 3, we compare the ratios of accretion in several star  
forming regions. 
Section 4 summarizes the results, stating the main conclusions.

%%%%%%%%%%%%%%%%%%%%%%%%%%%%%%%%%%%%%%%%%%%%%%%%%% 
%%%%%%%%%%%%%                   %%%%%%%%%%%%%%%%%% 
%%%%%%%%%%%%%    Section        %%%%%%%%%%%%%%%%%% 
%%%%%%%%%%%%%                   %%%%%%%%%%%%%%%%%% 
%%%%%%%%%%%%%%%%%%%%%%%%%%%%%%%%%%%%%%%%%%%%%%%%%% 

\section{H$\alpha$ equivalent width  as an empirical criterion of accretion and chromospheric activity}

\subsection{Infrared excesses and H$\alpha$  emission in T~Tauri stars} 
%%%%%%%%%%%%%%%%%%%%%%%%%%%%%%%%%%%%%%%%%%%%%%%%%%%%%%% 
 
Both classical and weak-line  T~Tauri 
stars have been identified using near infrared color-color diagrams. 
We have collected H$\alpha$ equivalent widths
 of T~Tauri stars 
belonging to different star forming regions and very young clusters 
and compared them with the color excesses. These color excesses 
were computed using the measured color indices and   
spectral type, and the typical color 
indices corresponding to the spectral type,  
 as listed by Bessell \& Brett (1988), Kirkpatrick et al$.$ (2000), 
and Leggett (1992) and Leggett et al$.$ (2000, 2002). 

Figure 1  illustrates one of  these comparisons for stars in Taurus. 
Open and solid circles denote the location of weak-line and classical 
T~Tauri stars, respectively. 
In this and other figures discussed below displaying W(H$\alpha$), we have 
avoided objects which have  
been classified as classical or weak-line T~Tauri stars or substellar 
analogs based on the W(H$\alpha$), i.e., the displayed CTTS and SCTTA  
show either broad H$\alpha$, forbidden lines, strong infrared excesses, or 
veiling.
All the data presented in this paper come from a large diversity of 
papers (see section 2.4 for references), and the corresponding 
spectra were collected at very different resolutions. This can translate 
into different values of the equivalent width for a given object (i.e.,  
lower resolution tends to yield larger W(H$\alpha$) than high resolution  
spectroscopy). This is a caveat that has to be kept in mind throught 
this and any other similar analysis. 

In Figure 1,  we display as short dashed lines two W(H$\alpha$) 
criteria used in the literature for classifying CTTS. 
Clearly, most of the CTTS are above the 20 \AA{} threshold, whereas 
most of the  WTTS have  W(H$\alpha$)$\le$5 \AA. 
However, there is an area in between where the transition from WTTS to CTTS 
 is not clear-cut. Part of the confusion arises from the fact that  
H$\alpha$ emission can also be produced by chromospheric activity. 
Therefore, our next step is to study the H$\alpha$ emission 
of active young stars. 

\subsection{The H$\alpha$ emission in young open clusters} 
%%%%%%%%%%%%%%%%%%%%%%%%%%%%%%%%%%%%%%%%%%%%%%%%%%%%%%% 
 
Figure 2a displays the ratio between the H$\alpha$ and the bolometric 
luminosities versus the spectral type for members of  
 three young open clusters, namely the Pleiades, Alpha Per and IC2391. 
The age ranges are 80--125 Myr, 50--90 Myr, and 30--53 Myr 
for each of them, depending  on the age dating technique:  
upper main-sequence isochrone 
fitting or lithium depletion at the substellar limit, respectively 
(Maeder \& Mermilliod 1981; Meynet et al$.$ 1993; 
 Stauffer et al$.$ 1998, 1999; Barrado y Navascu\'es et al$.$ 1999). 
The vertical, dotted lines delimit 
the age-dependent lithium gap in each cluster 
(see Barrado y Navascu\'es \& Stauffer 2003 and references therein).   
In order to convert  H$\alpha$ equivalent width into luminosity, 
 we followed a similar procedure as the one described 
by Mohanty \& Basri (2003). 
We computed fluxes corresponding to the continuum around 6563 \AA{ } 
--$F_{cont}$(H$\alpha$)-- 
using theoretical models by Allard et al$.$ (2001). 
We selected those having log~g=4.5, which correspond to 
ages between 50 and 100 Myr (Baraffe et al$.$ 1998). 
In particular, we used ``dusty'' models for T$_{eff}$$\le$3900 K and 
``NextGen'' models for higher effective temperatures. 
We note that ``NextGen'' models are good enough down to 
T$_{eff}$=2500 K (Mart\'{\i}n et al$.$ 2001b), but the predicted fluxes for 
 both sets are very similar and they do not affect the results.  
Then, the  flux of the  H$\alpha$ emission line was computed as  
$F_{line}$(H$\alpha$)= W(H$\alpha$)$\times$$F_{cont}$(H$\alpha$). 
Finally, the ratio can be expressed  as   
L(H$\alpha$)/L(bol) = $F_{line}$(H$\alpha$)/$\sigma$T$_{eff}$, where 
L(H$\alpha$) and L(bol) are the luminosity in the H$\alpha$ line and the bolometric  
luminosity. 
This diagram clearly shows a maximum at -3.3,  
which it is essentially defined by low mass stars with M2-M4 spectral types. 
 A similar diagram was 
obtained by Mohanty \& Basri (2003) for mid-M and L field objects  
(see their figure 7). 
 Besides H$\alpha$, this saturation limit also occurs in other  
activity indicators such as the X-ray luminosity  
(Stauffer et al$.$ 1994; Randich et al$.$ 1996) 
or in H$\beta$ (Delfosse et al$.$ 1998).
 
We have found that our saturation limit,  
log \{L(H$\alpha$)/L(bol)\}=$-$3.3, derived from the  
young open clusters IC2391, Alpha Per and the Pleiades, 
is somewhat higher than the values obtained by  
Delfosse et al$.$ (1998) and Mohanty \& Basri (2003) 
in field mid-dM stars, and late M and L spectral type  
very low mass stars stars and BDs.  
Since the first group (cluster members) are, on average, younger,  
and they have larger radius 
(the ratio between the radii of a 0.072 M$_\odot$ at 100 and 500 Myr is 4.7,  
according to Baraffe et al$.$ 1998 models),  
this difference suggests that there is an evolutionary component, 
perhaps related with size and/or surface gravity, 
in the saturation limit and, by extension, 
in the chromospheric activity in objects at the end of the dwarf  
sequence. A similar conclusion has been reached by Basri \& Mohanty (2003).

Now, we are interested on observational quantities, in particular  
W(H$\alpha$). Therefore, starting with this saturation limit, we have  
proceeded backward and derived the W(H$\alpha$) value for each spectral type 
which would correspond to it. 
Since we are trying to define a criterion for CTTS and SCTTA, 
which have maximum ages of about 10 Myr, we have used  
the atmosphere models corresponding to a gravity of log~g=4.0. 
Figure 2b shows the measured  W(H$\alpha$)  of members 
(low mass stars and BDs) of these three 
young clusters. 
 The bold--dotted line  corresponds to the 
 saturation limit at log \{L(H$\alpha$)/L(bol)\}=$-$3.3 dex. 
This criterion can be found in tabular form in Table 1. 
From now on, this criterion will be called the {\it saturation criterion}. 

The dashed line in Figure 2b corresponds to the 
upper envelope of the W(H$\alpha$) for the clusters. 
 For objects cooler than M5.5, the maximum W(H$\alpha$) measured is 20 \AA. 
Note, however, that the data are sparse at the low mass end of these
cluster sequences.
For stars warmer than this spectral type,  
the behavior is pseudo-linear in a logarithmic scale. 
 We have selected a fit which 
includes all the measurements. 
This curve, which is purely empirical, 
 will be called the young cluster {\it chromospheric criterion}  
from now on. 
The mathematical expression of this criterion is: 
 
\begin{equation} 
log W(H\alpha) = 0.0893 \times Sp.Type - 4.5767  
\end{equation} 
 
\noindent where the spectral type  O1 corresponds to 1, 
 B1 to 11, A1 to 21, etcetera. This fit is valid between  
spectral types G5 and M5.5. 
 
There is a clear tendency that the cooler the object, the larger  
the H$\alpha$ equivalent width, on average.  
The general reason for this trend is the decreasing photospheric
luminosity at the wavelength of H-alpha as the temperature decreases
(the so-called "contrast effect"; Basri \& Marcy 1995). The quantitative
behavior of the line strength, however, is not well understood.
We adopt this upper envelope as 
the maximum  emission due to chromospheric activity, 
without taking into account values measured during flares 
(known flares, such as that of PPl-15 
detected by Basri \& Mart\'{\i}n (1999), have been excluded).

\subsection{The H$\alpha$ emission in flare stars and field 
M and L objects} 
%%%%%%%%%%%%%%%%%%%%%%%%%%%%%%%%%%%%%%%%%%%%%%%%%%%%%%% 
 
Figure 3 is as Figure 2, but in this case a comparison  
with flare stars is displayed  (UV Cet type, crosses). 
The figure also includes cooler objects recently discovered  
by the 2MASS survey (Skrutskie et al$.$ 1997), 
 down to the L5 spectral type (plus symbols). 
Note the logarithmic scale in the y-axis. 
 
The selected UV Cet sample comes from Gershberg et al$.$ (1999). 
The IAU defines UV Cet variables as:  
``Dwarf stars of spectral classes dM3e-dM6e 
characterized by rare and very short flares 
 with amplitudes from 1 mag to 6 mag.  
Maximum brightness (usually sharp) is attained in a few, 
or several tens of seconds after the commencement 
 of the flare, total duration of the flare being equal 
 to about ten to fifty minutes''. 
A more general definition, which comes from 
 Gershberg et al$.$ (1999), is: 
``UV Cet-type variables are stars on the lower part of the main sequence  
which show phenomena inherent to the solar activity. The most  
manifestations of the solar activity are detected on such stars: 
 sporadic flares, dark spots, variable emissions from chromospheres  
and coronae, radio, X-ray and UV bursts.'' 
Therefore, this definition includes the first one 
(all the stars in the first group should be in the second). 
The sample shown in Figure 3 is based on this definition.

Regarding the cooler sample, the data come from Kirkpatrick et al$.$ (2000), 
Gizis et al$.$ (2000); Reid et al$.$ (2002); and Mohanty \& Basri (2003). 
In addition, stars and BDs with variable H$\alpha$, mainly due to the 
presence of flares, are also included as arrows and labeled. The beginning and the 
end of each  arrow   represent the minimum and maximum H$\alpha$ equivalent width 
ever recorded. The data come from 
Ruiz et al$.$ (1990); 
Irwin et al$.$ (1991);  
Eason et al$.$ (1992); 
Basri \& Marcy (1995),  
Kirkpatrick et al$.$ (1995);  
Mart\'{\i}n et al$.$ (1996, 1999, 2001),  
Delfosse et al$.$ (1997);  
Tinney et al$.$ (1998),
Tinney  (1999); 
Gizis et al$.$ (1999, 2000); 
Liebert et al$.$ (1999, 2003);  
Reid et al$.$ (1999, 2002);  
Hall (2002ab);  
Zapatero Osorio et al$.$ (2002b); and 
Mohanty et al$.$ (2003). 
%Alekseev et al$.$ (2001); 

This figure includes two different curves. 
The first one (bold, dotted line), also in Figure 2, 
is the saturation criterion. 
The second curve, plotted as a thin-solid line, is the upper envelope  
of the flare stars and field very low mass stars and  brown dwarfs, 
 which defines the {\it flare criterion}.

As the visual inspection and the comparison with Figure 2 show, most of the  
flare stars do have H$\alpha$ equivalent width which 
is less or equal than the  upper limit defined  
by cluster stars.  
%%%%%%%%%%%%% 
%Moreover, the saturation criterion  
%approximately  agrees with the maximum activity when flaring for objects 
% cooler than M4. Note the maximum in W(H$\alpha$) at M9, present  
%in both criteria. 
In any case, in order to establish the nature of an object 
with a strong H$\alpha$ emission, several consecutive spectra  
should be collected, to check the possibility of a flare. 
This an advisable strategy regardless the final goal, 
 when acquiring  spectra 
(i.e., splitting the total exposure time into three or four 
consecutive observations). 
%%%%%%%%%%%%%55 
%Note, moreover, that flares increase the equivalent width 
%of Balmer lines by a factor 1.5-3 in most cases (from quiescent 
%to maximum values), at least in the case of the early to mid-M dwarfs. 

\subsection{H$\alpha$ in very young stellar associations} 
%%%%%%%%%%%%%%%%%%%%%%%%%%%%%%%%%%%%%%%%%%%%%%%%%%%%%%% 
 
We have done the same exercise for several star forming regions 
 and very young clusters, 
with ages in the range 1-10 Myr, 
and displayed H$\alpha$ equivalent with against the spectral type. 
These results are depicted in Figure 4, which includes 
%%%%%%%%%%%% 
the Orion population  
(from  
 Herbig \& Bell 1988;  
 Alcal\'a et al$.$ 1996, 1998, 2000), 
%%%%%%%%%%%% 
Taurus  
(Brice\~no et al$.$ 1993, 1998, 1999, 2002;  
 Alencar \& Basri 2001; 
Mart\'{\i}n et al$.$ 2001a; 
 Luhman et al$.$ 2003a; 
 White \& Basri 2003;   
 Muzerolle et al$.$ 2003) 
%%%%%%%%%%%% 
IC348  
(Herbig 1998; 
 Luhman et al$.$ 1999, 2003b;  
 Muzerolle et al$.$ 2003; 
 Jayawardhana et al$.$ 2003a), 
%%%%%%%%%%%% 
Sco-Cen-Lupus-Crux Complex, including the $\rho$ Oph molecular cloud 
(Bouvier \& Appenzeller 1992;  
 Mart\'{\i}n et al$.$ 1998, 2004, in prep; 
 Ardila et al$.$ 2000; 
 Jayawardhana et al$.$ 2002a;  
 Mamajek et al$.$ 2002),  
%%%%%%%%%%%% 
Chamaeleon~I  
(Guenther et al$.$ 1997; 
 Comer\'on et al$.$ 2000; 
 Saffe et al$.$ 2003;  
 Jayawardhana et al$.$ 2003),  
%%%%%%%%%%%% 
$\sigma$ Orionis cluster  
(B\'ejar et al$.$ 1999; 
 Barrado y Navascu\'es et al$.$ 2001, 2002, 2003: 
 Zapatero Osorio et al$.$ 2002a), and 
%%%%%%%%%%%% 
TW~Hydra association --TWA--  
(Sterzik et al$.$ 1999; 
 Webb et al$.$ 1999; 
 Zuckerman et al$.$ 2001; 
 Gizis  2002; 
 Mohanty et al$.$ 2003). 

 In these figures,  solid circles, small open circles and  
 open triangles correspond to  
CTTS/SCTTA,   WTTS/SWTTA  and  PTTS, respectively.  
These classifications are based on criteria that do not use 
the W(H$\alpha$). 
Small crosses denote unclassified objects.
Note the logarithmic scale in the y-axis. 
Stars and BDs with mid-infrared excesses (signpost of  
circumstellar disks)  are indicated as 
large open circles. They were selected from 
Natta \& Testi (2001), Natta et al$.$ (2002), Testi et al$.$ (2002),
 Comer\'on   et al$.$ (1998), 
  and Jayawardhana et al$.$ (2002b, 2003b). 
In the case of the $\sigma$ Orionis cluster, the large solid and  
broken circles denote the location of cluster members with near infrared excesses 
(Barrado y Navascu\'es et al$.$ 2003). 
Large open squares correspond to objects with forbidden lines  
(Zapatero Osorio et al$.$ 2002a; Muzerolle et al$.$ 2003; Brice\~no et al$.$ 1998; 
Barrado y Navascu\'es et al$.$ 2004, in prep.), 
which characterize outflows. 
The vertical dotted segment is located at the spectral type which  
divides the stellar from the substellar domain (for ages younger than $\sim$100 Myr). 
 The bold, dotted curve corresponds to the saturation 
criterion to classify CTTSs and substellar analogs. 
Traditional criteria dividing between classical and weak-line 
T~Tauri stars, such as the 5 or 20 \AA{ } limits (long-dashed  
horizontal lines), are also included. 
Note that these criteria do not describe the CTTS/WTTS phenomenology 
appropriately. The criterion proposed by Mart\'{\i}n (1998). 
which depends on the spectral type, is more reliable. Our saturation criterion 
improves upon the previous ones and extends 
into the substellar domain.

\subsection{A new empirical criterion to classify classical T~Tauri 
stars and substellar analogs} 
 
When weighting together all the data and criteria presented in the  
sections above, we have noted that most CTTS  have H$\alpha$ 
equivalent widths larger than the characteristic values of  
older stars (i.e., cluster and field stars) even when these last ones are  
flaring. Our main goal is to define an empirical  
spectroscopic criterion to  
classify any object as accreting or non-accreting    
using solely low-resolution optical spectroscopy. 
The data clearly indicate that all stars and brown  
dwarfs, except the few exceptions --mostly having large variability--,  
that are   located above the saturation criteria 
(the H$\alpha$ equivalent width corresponding to log \{L(H$\alpha$)/L(bol)\}=$-$3.3) 
are, indeed, accreting and, therefore, they can be classified as 
bona-fide  CTTS and SCTTA. 
This is a rather restricting  criterion: 
all objects above this line should be CTTS and SCTTA  
(except for very strong peaks of flare activity which  
are short lived and rare). Below this dividing line 
 there could be objects with a low-level of accretion.   
Furthermore, we note that  
throughout this paper, we have used the measured equivalent widths,  
without any correction due to the veiling of the continuum. Therefore,  
true equivalent widths should be equal or larger for CTTS  
 and SCTTAs than the measured values,  
and all or some of the accreting  objects located below the saturation criterion 
might be, in fact, above it once the veiling correction is performed. 
Unfortunately, the data in the literature are incomplete and very inhomogeneous 
regarding the technique to derive the veiling, and we have opted  
for using the uncorrected values. 
 
In order to make easier the visual inspection, and to compare our criterion with 
all the available data, we have included  in Figure 5  
 the  information displayed in Figures 4a-4h. 
Figure 5 also shows (solid line) the re-definition  
of the criterion proposed by Mart\'{\i}n (1998),  
carried out by White \& Basri (2003). 
Our proposed criterion has a sound physical  
basis: the maximum amount of energy which can be released in non-thermal  
processes by the chromosphere, $\sim$5$\times$10$^{-4}$ of the total energy. 
 Objects exceeding this limit  must have drawn it from other sources, such as accretion 
from a disk. 
This criterion  is listed in Table 1, which contains the H$\alpha$ equivalent  
width depending on the spectral type.

As a conclusion, a moderate signal-to-noise (S/N$\sim$35),  
low resolution (R$\sim$600) spectrum  
is enough to detect most of the CTTS and SCTTA. 
There are exceptions, such as a post-T~Tauri star 
belonging to the Sco-Cen complex, 
with W(H$\alpha$) larger  than the saturation criterion.  
However,  additional information, such as the lithium abundance  
measured in the stellar photosphere, is enough to disentangle this ambiguity  
(see Mart\'{\i}n et al$.$ 1997, 1998). 
Finally, some  of the substellar objects with disks 
(detected via their infrared excesses 
or because of the presence of forbidden lines in their spectrum) do not satisfy  
the proposed saturation criterion. We conclude that the amount of material  
they are accreting should be relatively small when compared with more massive 
accretors. This might suggest, too, that the time scale for disk dissipation  
is smaller in brown dwarfs. 
The veiling might be affecting by a large amount their evolution, when compared with  
accreting stars. 
 Therefore, besides the saturation criterion, 
additional information is very helpful 
to detect accretion in most of the substellar objects, 
such as W(He{\sc I}5876), W(He{\sc I}6678) or the ratios between the 
components of the CaII infrared triplet. 
Note, however, that they have H$\alpha$ above the cluster chromospheric criterion, 
which equals 20 \AA{ } in this region (objects cooler than M5.5).

%%%%%%%%%%%%%%%%%%%%%%%%%%%%%%%%%%%%%%%%%%%%%%%%%% 
%%%%%%%%%%%%%                   %%%%%%%%%%%%%%%%%% 
%%%%%%%%%%%%%    Section        %%%%%%%%%%%%%%%%%% 
%%%%%%%%%%%%%                   %%%%%%%%%%%%%%%%%% 
%%%%%%%%%%%%%%%%%%%%%%%%%%%%%%%%%%%%%%%%%%%%%%%%%% 
 
\section{Accretion frequency} 
 
We have made use of our empirical criterion 
--the saturation at log \{L(H$\alpha$)/L(bol)\}=$-$3.3-- 
to determine the number of CTTS and substellar  analogs in several 
star forming regions, and to compute the fraction of accretors 
(i.e., the fraction of objects harboring an active circumstellar 
or circum-substellar disk). 
Table 2 lists the name of the association or cluster, their age 
(including maximum and minimum values commonly quoted 
in the literature) and the fraction of accreting objects, 
 as computed for several spectral ranges. 
 Since the saturation criterion  
--based on the equivalent width of H$\alpha$-- is very restrictive,  
the values compiled in the table can be considered as  
lower limits. 
Figure 6 displays, in a log-log plane, this information. 
 In this figure, we depict 
in three different panels the fraction for stellar members (K3-M5.5) 
and substellar components (M5.5-L2 and M5.5-M7.5). 
The selection of the M5.5-M7.5 range responds to the need   of
 avoiding biases due  to incompleteness of the surveys. However, we do not  
appreciate any significant difference. 
We have selected the spectral type M5.5 as the substellar borderline  
based on  theoretical models --1 or 10 Myr isochrones-- 
 by Baraffe et al$.$ (1998). 
 
Although there are significant uncertainties due to poor statistics 
in some cases,  ages derived in heterogeneous ways 
(different evolutionary tracks or methodology to  
convert theoretical parameters into observational quantities, see 
Stauffer et al$.$ 1995), age spread within the association, an so on, 
 the decline of the disk frequency seems to be  linear. 

The data suggest that the accretion  
disk life-times are similar in the substellar domain than for more  
massive objects. We note the low fraction of accreting stars 
found in Rho Oph, although this could be related to strong veiling which  
could diminish the measured H$\alpha$ equivalent widths and, hence, the 
fraction of accretors estimated from H$\alpha$. 
Recent works (Muench et al. 2001, Haisch et al. 2001, 
Jayawardhana et al. 2003b, Liu et al. 2003) have derived the fraction of objects 
with disks using different techniques for these and other SFR. 
Although the ratios are systematically larger in those studies 
-it has been stated before that our fraction represents only a lower limit--, 
the trend of decreasing disk emission with age is similar to what we find. 
Thus, W(H$\alpha$) alone is adequate tool to study the frequency of accreting 
objects in clusters and star-forming regions.  

%%%%%%%%%%%%%%%%%%%%%%%%%%%%%%%%%%%%%%%%%%%%%%%%%% 
%%%%%%%%%%%%%                   %%%%%%%%%%%%%%%%%% 
%%%%%%%%%%%%%    Section        %%%%%%%%%%%%%%%%%% 
%%%%%%%%%%%%%                   %%%%%%%%%%%%%%%%%% 
%%%%%%%%%%%%%%%%%%%%%%%%%%%%%%%%%%%%%%%%%%%%%%%%%% 
 
\section{Final remarks} 
 
Based primarily on the saturation limit at log \{L(H$\alpha$)/L(bol)\}=$-$3.3,  
we have been able  
to establish an empirical criterion to distinguish between     
CTTS and WTTS, 
and we have extended it into the substellar domain. An 
schematic classification is shown in Figure 7. 
Four regions are defined in this figure. Region A is populated by 
CTTSs and their substellar analogs. They  have 
 W(H$\alpha$) above the line defined by the saturation  
(i.e., the origin cannot be chromospheric).  Since the equivalent width values are very  
high, a moderate signal-to-noise, low resolution optical spectrum  
is enough to detect the presence of accretion. 
In region (B) classical, weak-line T~Tauri and flaring objects can be found. 
Most of the objects in this area should be BDs. 
In particular, all WTTS and CTTS with spectral type cooler than M5.5-M6. 
This area is delimited by the saturation criterion (up) and  
the maximum chromospheric activity measured in upper clusters (down).

Since there is a mixture of objects in region B, it is useful to obtain additional 
information about the objects. \\  
b1.- Several consecutive spectra at moderate S/N, low resolution, 
can establish whether the object is flaring, since the H$\alpha$ 
equivalent widths change very fast and with a very characteristic pattern.\\ 
b2.- Spectroscopy at R$\sim$3000 resolution 
 ($\Delta$v=100 km/s) allows a rough study of the  
H$\alpha$ line profile and width. 
In the case of very low mass 
stars and brown dwarfs, White \& Basri (2003) have  proposed FWHM=270 km/s as  
a criterion to establish the presence of accretion (i.e., CTTS/SCTTA).\\ 
b3.- At this resolution,  
the lithium feature at 6708 \AA{ } can be detected (S/N$\sim$35). 
The lack of lithium denotes that the object  
is older than the time needed for disk dissipation ($\sim$10 Myr) 
and must have a mass larger than 0.060 M$_\odot$.\\ 
b4.- Other lines and line  ratios, measured in low or medium resolution spectra, 
 such as those  He{\sc i}5876, He{\sc i}6677 and  Ca{\sc ii} IRT, can differentiate between  
CTTS/SCTTA, WTTS/SWTTA, and older objects experiencing flares. 

In region C there are classical, weak-line and post-T~Tauri stars. 
The limits are the same as for region B, but in this case, this region  is located at  
the warmer  side of the diagram. Additional information that helps to classify 
objects in this regions includes spectroscopy with resolution R$\sim$3000, which  is enough to  
  measure the width of the H$\alpha$ line 
and use the   10\% FWHM criterion, which allows the distinction between  
SCTTA and SWTTA. 
The lithium  depletion or absence of it is an adequate tool for differentiating between 
T~Tauri and post-T~Tauri stars.
 
Region D is populated by cluster and field stars and BDs.
This area also includes weak-line and  post-T~Tauri objects and is delimited  
by the upper envelope of the activity measured in young open clusters. 
Additional criteria to classify objects in region D include:\\ 
d1.- As in b1, several low resolution spectra are essential to detect  
a flare.\\ 
d2.- Again, a  R$\sim$3000 spectrum suffices to detect the lithium feature 
in an object with a spectral type later than about M0, 
and to measure the activity. Lithium equivalent width can be compared with the values  
shown by cluster members, weak-line and post-T~Tauri stars and brown dwarfs.\\ 
d3.- Moderate resolution spectroscopy (R$\ge$10,000) provides  
measurements of the rotational velocity ($vsini$). Since there is  
a relationship between age and rotation (slower rotators being older for a given spectral type,  
Stauffer et al$.$ 1987ab, 1989) for dwarfs earlier than about M6, 
this information characterizes, from the statistical point of view, 
the evolutionary status.\\ 
d4.- Finally, lithium equivalent widths can be measured in  
objects with spectral type earlier than M0 if a spectrum with resolution  
larger than R=10,000 is available.  
 
In all these four cases, the detection or lack of forbidden lines,  
as well as other permitted lines, as discussed in section 4.1,  
can put strong constrains on the CTTS/SCTTA or WTTS/SWTTA nature. 
 
As a conclusion, by re-analyzing H$\alpha$ 
 and lithium equivalent  
widths in members of several young open clusters and 
star forming regions, we have been able to define well-defined 
spectroscopic  criterion based on the saturation limit for  
young open clusters, which can be used to statistically classify 
populations of stars and BDs in young open clusters and 
star-forming regions using low-resolution optical spectroscopy, 
which is efficiently obtained using multi-object spectrographs. 
 
\begin{acknowledgements} 
 Financial support was provided by the Spanish 
{\it ``Programa Ram\'on y Cajal''}  
and AYA2001-1124-CO2 programs. 
Partial funding was provided by the National Aeronautics and Space 
Administration (NASA) grant NAG5-9992 and  
National Science Foundation (NSF) grant AST-0205862. 
 We are truly indebted to the referee, Gibor Basri, for his very
useful comments.
\end{acknowledgements}

%%%%%%%%%%%%%%%%%%%%%%%%%%%%%%%%%%%%%%%%%%%%%%%%%% 
%%%%%%%%%%%%%                   %%%%%%%%%%%%%%%%%% 
%%%%%%%%%%%%%    Section        %%%%%%%%%%%%%%%%%% 
%%%%%%%%%%%%%                   %%%%%%%%%%%%%%%%%% 
%%%%%%%%%%%%%%%%%%%%%%%%%%%%%%%%%%%%%%%%%%%%%%%%%% 

\clearpage 
%  %  %  %  %  %  %  %  %  %  %  %  %  %  %  %  %  %  %  %  %  %  %  %  %  %  %  %  %  %  %  %  %  %  %  %   
%  %  %  %  %  %  %  %  %  %  %  %  %  %  %  %  %  %  %  %  %  %  %  %  %  %  %  %  %  %  %  %  %  %  %  %   
%  %  %  %  %  %  %  %  %  %  %  %  %  %  %  %  %  %  %  %  %  %  %  %  %  %  %  %  %  %  %  %  %  %  %  %   

\begin{table} 
\caption[]{CTTS criterion, expressed in equivalent widths, 
 as derived from the saturation limit  
at  log~\{L(H$\alpha$)/L(bol)\}=-3.3 } 
\begin{tabular}{lr} 
\hline 
Sp.Type &  W(H$\alpha$) \\ 
        &  (\AA)        \\ 
\hline 
\hline 
  K0 &  3.9  \\ 
  K1 &  3.9  \\ 
  K2 &  4.0  \\ 
  K3 &  4.1  \\ 
  K4 &  4.4  \\ 
  K5 &  5.1  \\ 
  K6 &  5.9  \\ 
  K7 &  6.6  \\ 
  K8 &  7.2  \\ 
  K9 &  7.8  \\ 
  M0 &  8.7  \\ 
  M1 & 10.1  \\ 
  M2 & 11.2  \\ 
  M3 & 12.2  \\ 
  M4 & 14.7  \\ 
  M5 & 18.0  \\ 
  M6 & 24.1  \\ 
  M7 & 41.9  \\ 
  M8 & 53.0  \\ 
  M9 & 87.9  \\ 
  L0 &148.2  \\ 
  L1 &190.4  \\ 
  L2 &279.6  \\ 
  L3 &328.9  \\ 
  L4 &436.4  \\ 
  L5 &698.3  \\ 
%  L6 &726.1  \\ 
\hline 
\end{tabular} 
\end{table}

\begin{table*} 
\small 
\tiny 
\caption[]{Fraction of  CTT stars and substellar analogs, in percentage.
 Inside the parenthesis, we  list the number of accretors versus
 the total number of members ($N/total$). Errors have been computed as $\sqrt[2]{N}$/$total$.} 
\begin{tabular}{lccccccccc} 
\hline   
SFR &\multicolumn{3}{c}{Age}&$\;$&        K3--M5.5             &$\;$&          M5.5--L2         &$\;$&      M5.5--M75            \\ 
     \cline{2-4}                         \cline{6-6} 	                      \cline{8-8}               \cline{10-10} 
 &\multicolumn{3}{c}{Min. Ave. Max}&$\;$&                      &    &                           &    &                           \\ 
    &\multicolumn{3}{c}{(Myr)}     &$\;$&    (\%  )              &$\;$&            (\%  )           &$\;$&     (\%  )                  \\  
\hline				     		   		          
\hline				     		   		        
Orion  & 0.5 &  1   & 1.5 &$\;$&  68.3  $\pm$ 5.6 (151/221)  &$\;$& 50.0 \hspace{1cm} (1/2)    &$\;$& 50.0  \hspace{1cm}(1/2)   \\ 
RhoOph & 0.5 &  1   & 1.5 &$\;$&  24.4  $\pm$ 5.2  (22/90)   &$\;$& 20.0  $\pm$ 19.99 (1/5)    &$\;$& 50.0  \hspace{1cm}(1/2)\\ 
Taurus & 1   &  1.5 & 2   &$\;$&  52.9  $\pm$ 6.1  (93/158)  &$\;$& 50.0  $\pm$ 12.9 (15/30)   &$\;$& 45.5   $\pm$  14.4 (10/22)\\ 
IC348  & 1   &  2   & 3   &$\;$&  30.6  $\pm$ 4.4  (48/157)  &$\;$& 40.0  $\pm$  8.9 (20/50)   &$\;$& 41.5   $\pm$  10.1 (17/41)\\ 
ChaI   & 1   &  2   & 3   &$\;$&  53.3  $\pm$13.3  (16/30)   &$\;$& 25.0  $\pm$ 12.5  (4/16)   &$\;$& 28.6   $\pm$  14.3  (4/14)\\ 
SOri   & 2   &  5   & 8   &$\;$&  25.0  $\pm$ 9.4   (7/28)   &$\;$& 13.9  $\pm$  6.2  (5/36)   &$\;$& 14.3   $\pm$   8.2  (3/21)\\ 
UpSco  & 3   &  5   &12   &$\;$&  14.4  $\pm$ 5.7   (4/35)   &$\;$& 16.3  $\pm$  6.2  (7/43)   &$\;$& 13.5   $\pm$   6.0  (5/37)\\ 
TWA    & 5   & 10   &15   &$\;$&  10.7  $\pm$ 6.1   (3/28)   &$\;$&  7.1  $\pm$  7.09 (1/14)   &$\;$&  0.0  \hspace{1cm}(0/6)   \\ 
\hline  
\end{tabular} 
\end{table*} 
 
\clearpage 
%  %  %  %  %  %  %  %  %  %  %  %  %  %  %  %  %  %  %  %  %  %  % 
%  %  %  %  %  %  %  %  %  %  %  %  %  %  %  %  %  %  %  %  %  %  % 
%  %  %  %  %  %  %  %  %  %  %  %  %  %  %  %  %  %  %  %  %  %  % 

\setcounter{figure}{0} 
%  ----------------------------------------------------------- 
   \begin{figure*} 
   \centering 
   \includegraphics[width=14.0cm]{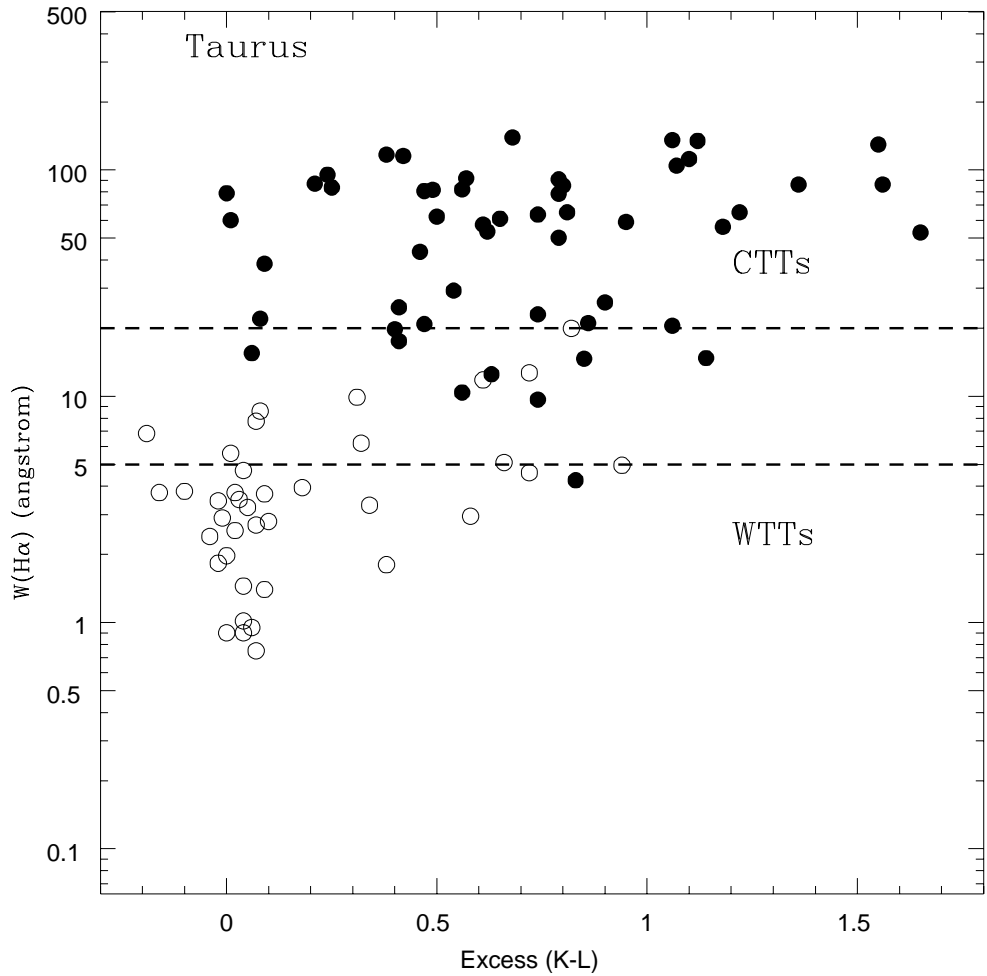} 
\caption{Equivalent width of H$\alpha$ --logarithmic scale--  against the color excess 
for the Taurus population (after Kenyon \& Hartmann 1995). 
Classical and weak-line T~Tauri are shown as solid and open circles, 
respectively. 
The two dashed lines delimit the areas for these two types 
of objects, based on the H$\alpha$ emission.  
} 
         \label{} 
   \end{figure*} 
%  ______________________________________________________________       
 
\clearpage 
%  %  %  %  %  %  %  %  %  %  %  %  %  %  %  %  %  %  %  %  %  %  %  %  %  % 
%  %  %  %  %  %  %  %  %  %  %  %  %  %  %  %  %  %  %  %  %  %  %  %  %  % 
%  %  %  %  %  %  %  %  %  %  %  %  %  %  %  %  %  %  %  %  %  %  %  %  %  % 

\setcounter{figure}{1} 
%  ----------------------------------------------------------- 
   \begin{figure*} 
   \centering 
   \includegraphics[width=14.0cm]{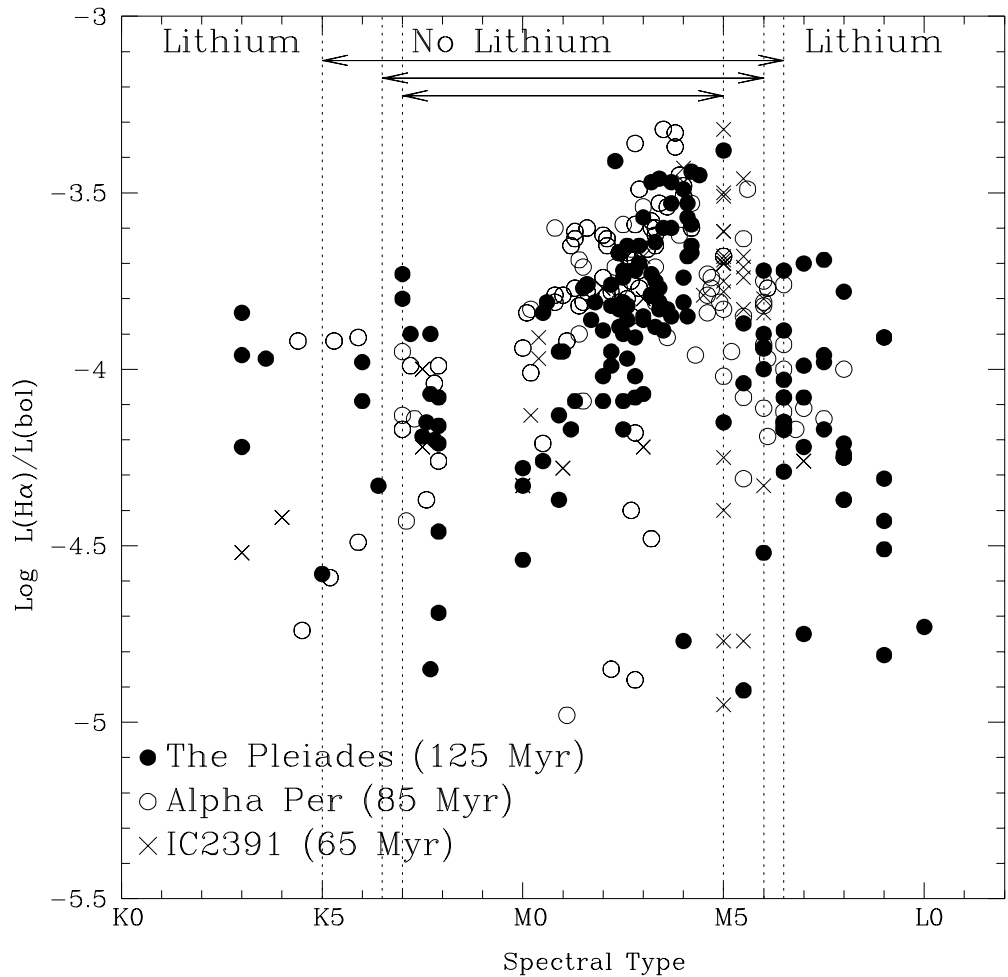} 
\caption{ 
{\bf a} Ratio between  the H$\alpha$ and the bolometric luminosities  
versus the spectral type for several young 
open clusters: 
the Pleiades (80-120 Myr), Alpha Per (50-90 Myr), 
and IC2391 (30-55 Myr).  
The vertical, dotted lines delimit  
the location of of the spectral type ranges where the 
lithium has been preserved or depleted. 
{\bf b} 
Same as panel a,  H$\alpha$ equivalent widths are  
plotted in the y-axis. 
 The dashed segments  correspond to the 
upper envelope of the W(H$\alpha$) for these three clusters 
(i.e., the maximum emission due to chromospheric activity). 
The bold, dotted curve is the saturation limit 
at  log \{L(H$\alpha$)/L(bol)\}=-3.3. 
Note the linear scale in the y-axis. 
} 
         \label{} 
   \end{figure*} 
%  ______________________________________________________________       
 
\clearpage 
%  %  %  %  %  %  %  %  %  %  %  %  %  %  %  %  %  %  %  %  %  %  %  %  %  % 
%  %  %  %  %  %  %  %  %  %  %  %  %  %  %  %  %  %  %  %  %  %  %  %  %  % 
%  %  %  %  %  %  %  %  %  %  %  %  %  %  %  %  %  %  %  %  %  %  %  %  %  % 
 
\setcounter{figure}{1} 
%  ----------------------------------------------------------- 
   \begin{figure*} 
   \centering 
   \includegraphics[width=14.0cm]{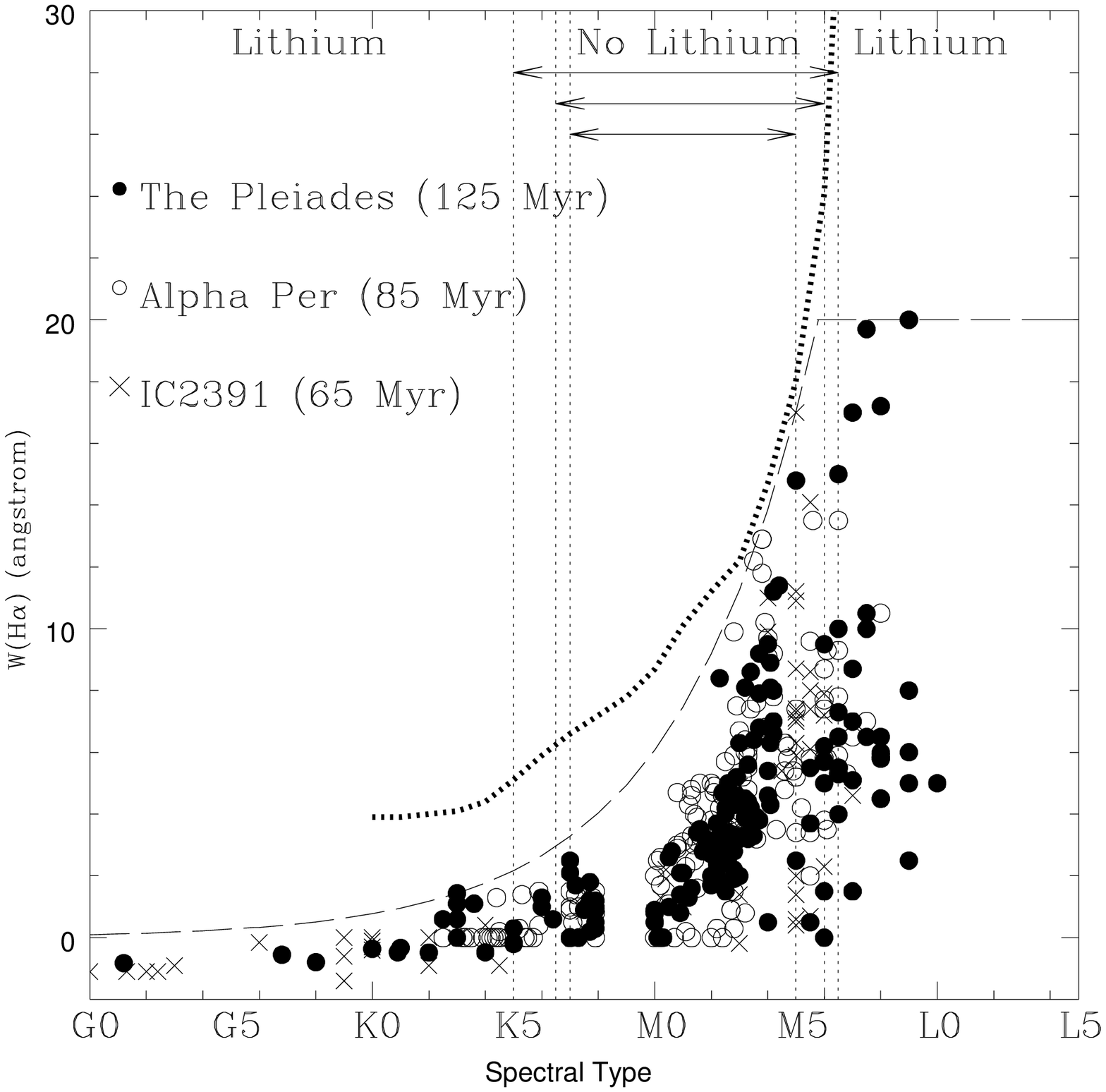} 
\caption{ 
(continue) 
} 
         \label{} 
   \end{figure*} 
%  ______________________________________________________________       
 
\clearpage 
%  %  %  %  %  %  %  %  %  %  %  %  %  %  %  %  %  %  %  %  %  %  %  %  %  % 
%  %  %  %  %  %  %  %  %  %  %  %  %  %  %  %  %  %  %  %  %  %  %  %  %  % 
%  %  %  %  %  %  %  %  %  %  %  %  %  %  %  %  %  %  %  %  %  %  %  %  %  % 

\setcounter{figure}{2} 
%  ----------------------------------------------------------- 
   \begin{figure*} 
   \centering 
   \includegraphics[width=14.0cm]{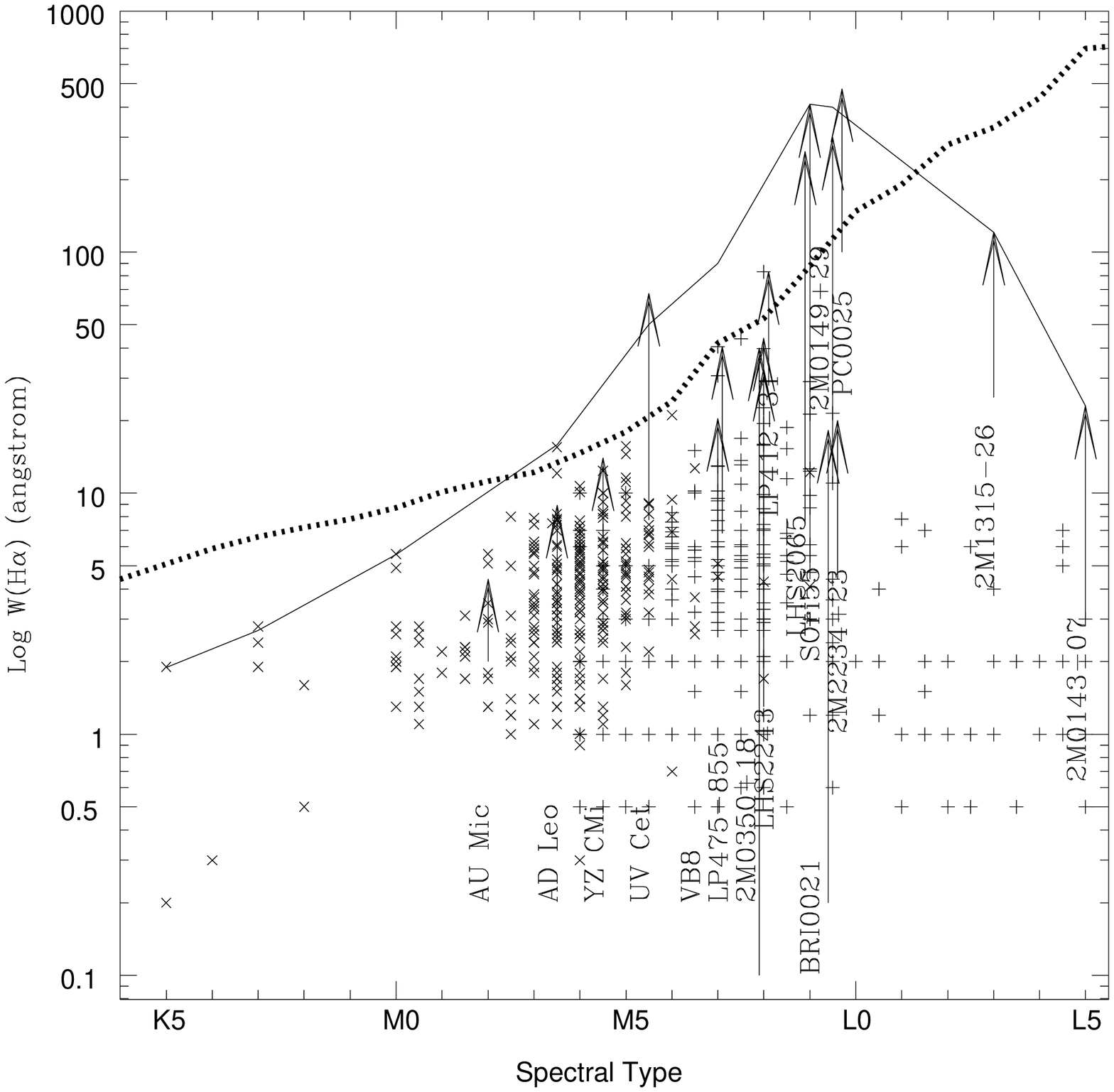} 
\caption{H$\alpha$ equivalent widths for  
UV Cet stars (crosses, after Gershberg et al$.$ 1999) 
 and M and L objects, both very low mass stars and BDs, from 2MASS 
(plus signs, from Kirkpatrick et al$.$ 2000; Gizis et al$.$ 2002; 
Reid et al$.$ 2002; Mohanty et al$.$ 2003). 
 Data corresponding to flares of several 
very active  stars are also displayed as arrows. 
 The  thin solid  curve corresponds to the 
upper envelope of flare stars and field object 
(i.e., the maximum emission for the flare of variable objects). 
The dotted, thick curve is the saturation limit. 
Note that most late M and L objects only have W(H$\alpha$) upper limits.} 
         \label{} 
   \end{figure*} 
%  ______________________________________________________________       
 
\clearpage 
%  %  %  %  %  %  %  %  %  %  %  %  %  %  %  %  %  %  %  %  %  %  %  %  %  % 
%  %  %  %  %  %  %  %  %  %  %  %  %  %  %  %  %  %  %  %  %  %  %  %  %  % 
%  %  %  %  %  %  %  %  %  %  %  %  %  %  %  %  %  %  %  %  %  %  %  %  %  % 

\setcounter{figure}{3} 
%  ----------------------------------------------------------- 
   \begin{figure*} 
   \centering 
   \includegraphics[width=14.0cm]{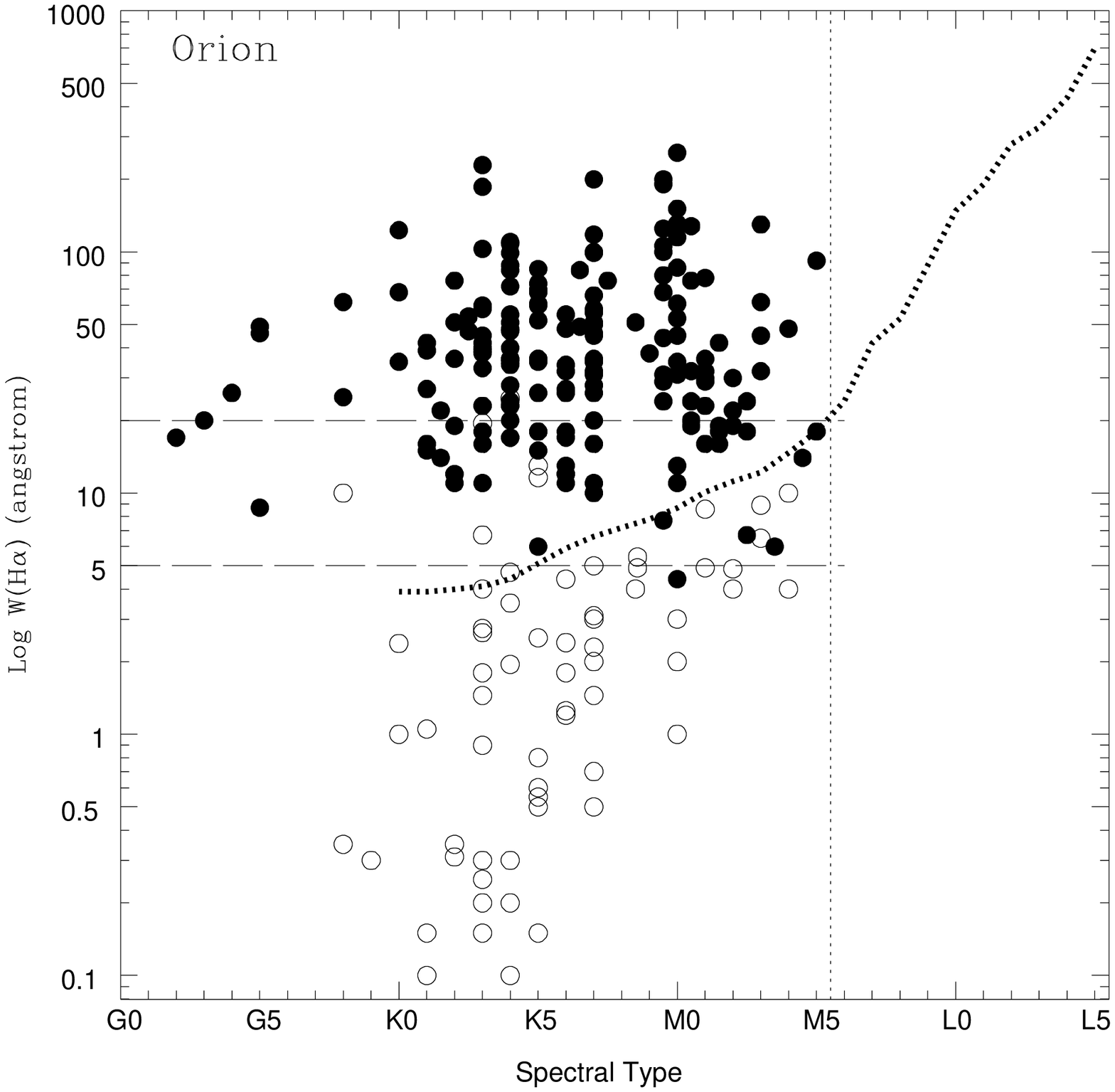} 
\caption{ H$\alpha$ equivalent widths for several 
star forming regions: 
Orion population, Rho Oph, Taurus, 
IC348, ChaI, $\sigma$ Orionis cluster, UpperSco, 
and the TW Hya association (see text for sources). 
Solid circles correspond to CTTS,  
open circles represent WTT stars and 
open triangles locate the position in the diagram of post-T~Tauri.   
Objects with no classification are shown as crosses. 
Large open circles represent objects with mid-IR excesses. 
In the case of the  $\sigma$ Orionis cluster, the same symbols depict  
objects with near IR excesses (broken circles have possible excesses). 
Finally,  large open squares denote objects with forbidden lines  
in their spectrum.  
The dotted, bold curve is the saturation  criterion, 
whereas two previously proposed criteria (5 and 20 \AA) to separate  
CTTS and WTTS (see section 3.4) 
are included as long-dashed, thin horizontal segments. 
The vertical dotted segment denotes the location of the substellar  
frontier for 100 Myr (i.e., longer than the expected lifetime 
of any circumstellar disk).  
} 
         \label{} 
   \end{figure*} 
%  ______________________________________________________________       
 
\clearpage 
%  %  %  %  %  %  %  %  %  %  %  %  %  %  %  %  %  %  %  %  %  %  %  %  %  % 
%  %  %  %  %  %  %  %  %  %  %  %  %  %  %  %  %  %  %  %  %  %  %  %  %  % 
%  %  %  %  %  %  %  %  %  %  %  %  %  %  %  %  %  %  %  %  %  %  %  %  %  % 
 
\setcounter{figure}{3} 
%  ----------------------------------------------------------- 
   \begin{figure*} 
   \centering 
   \includegraphics[width=14.0cm]{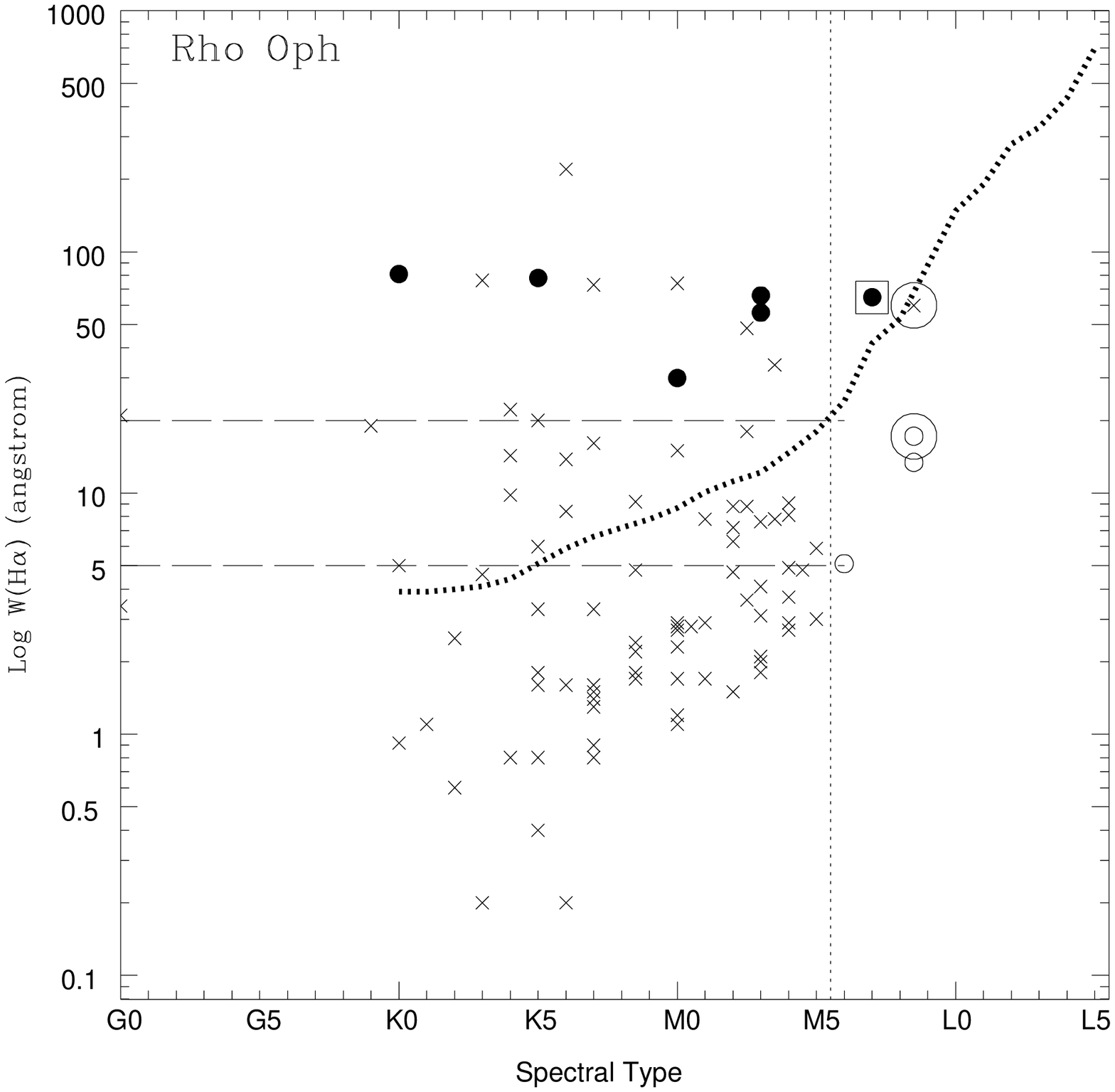} 
\caption{ (continue)} 
         \label{} 
   \end{figure*} 
%  ______________________________________________________________       
 
\clearpage 
%  %  %  %  %  %  %  %  %  %  %  %  %  %  %  %  %  %  %  %  %  %  %  %  %  % 
%  %  %  %  %  %  %  %  %  %  %  %  %  %  %  %  %  %  %  %  %  %  %  %  %  % 
%  %  %  %  %  %  %  %  %  %  %  %  %  %  %  %  %  %  %  %  %  %  %  %  %  % 
 
\setcounter{figure}{3} 
%  ----------------------------------------------------------- 
   \begin{figure*} 
   \centering 
   \includegraphics[width=14.0cm]{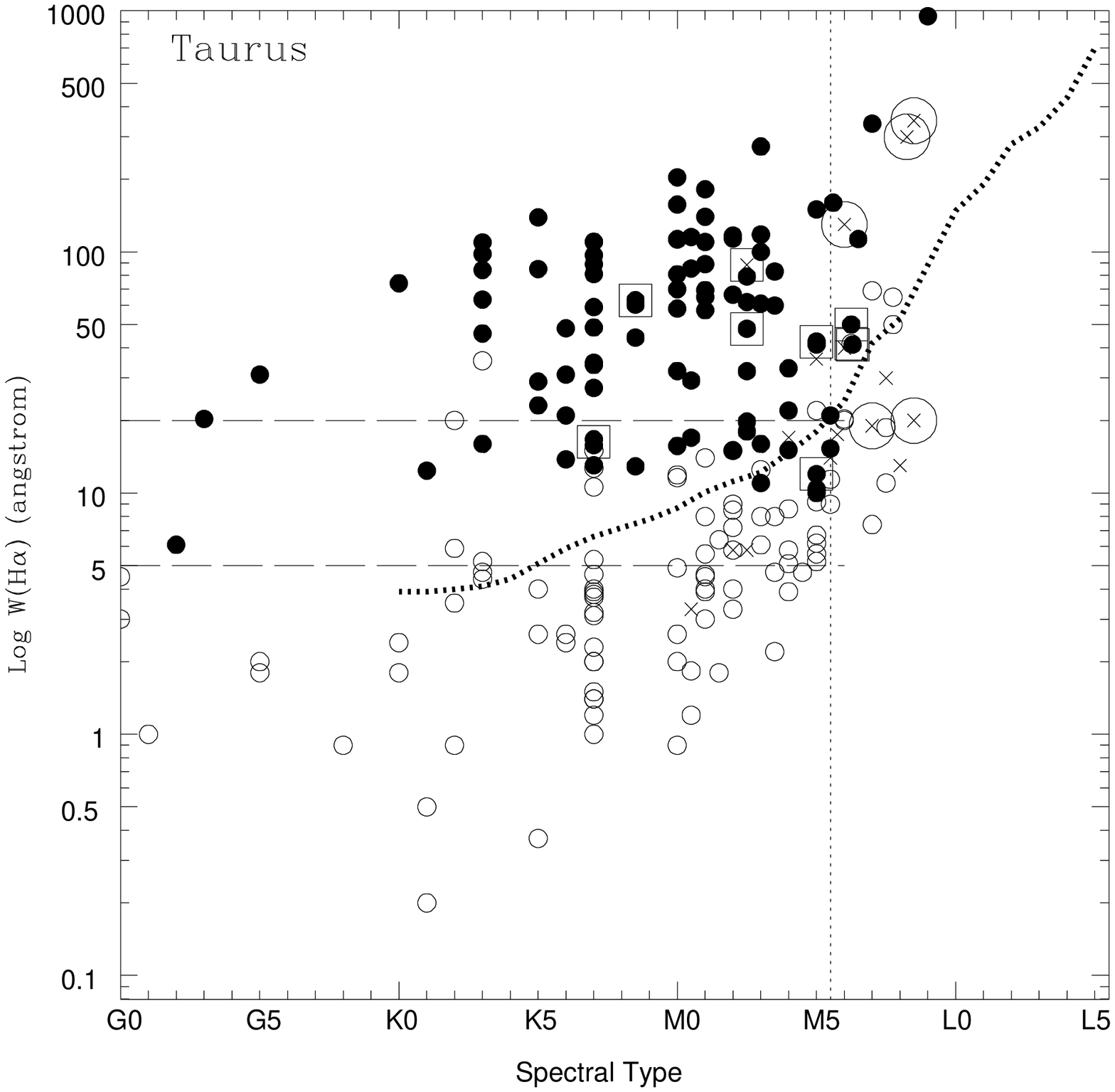} 
\caption{ (continue) 
} 
         \label{} 
   \end{figure*} 
%  ______________________________________________________________       
 
\clearpage 
%  %  %  %  %  %  %  %  %  %  %  %  %  %  %  %  %  %  %  %  %  %  %  %  %  % 
%  %  %  %  %  %  %  %  %  %  %  %  %  %  %  %  %  %  %  %  %  %  %  %  %  % 
%  %  %  %  %  %  %  %  %  %  %  %  %  %  %  %  %  %  %  %  %  %  %  %  %  % 
 
\setcounter{figure}{3} 
%  ----------------------------------------------------------- 
   \begin{figure*} 
   \centering 
   \includegraphics[width=14.0cm]{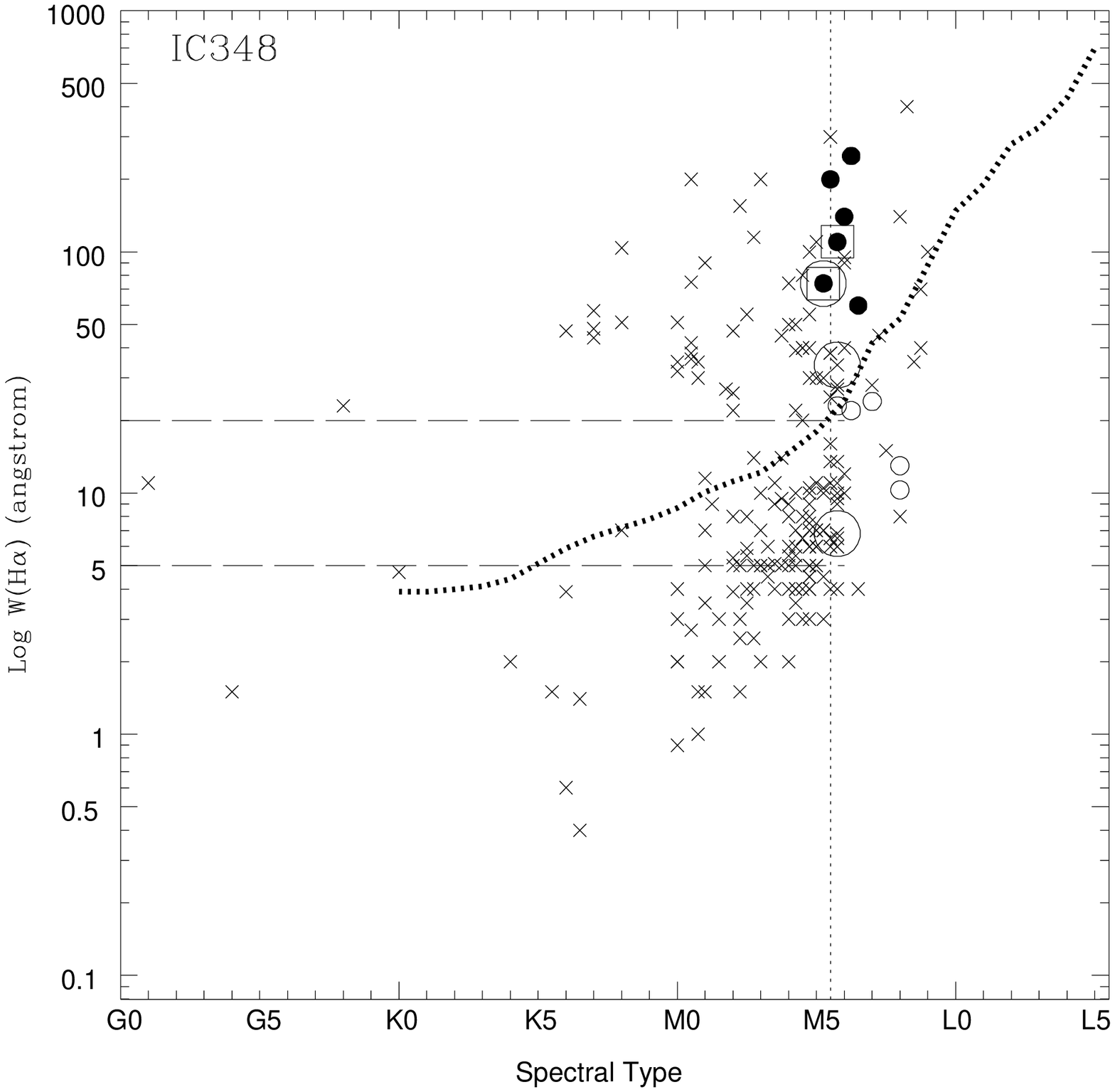} 
\caption{ (continue)  
} 
         \label{} 
   \end{figure*} 
%  ______________________________________________________________       
 
\clearpage 
%  %  %  %  %  %  %  %  %  %  %  %  %  %  %  %  %  %  %  %  %  %  %  %  %  % 
%  %  %  %  %  %  %  %  %  %  %  %  %  %  %  %  %  %  %  %  %  %  %  %  %  % 
%  %  %  %  %  %  %  %  %  %  %  %  %  %  %  %  %  %  %  %  %  %  %  %  %  % 
 
\setcounter{figure}{3} 
%  ----------------------------------------------------------- 
   \begin{figure*} 
   \centering 
   \includegraphics[width=14.0cm]{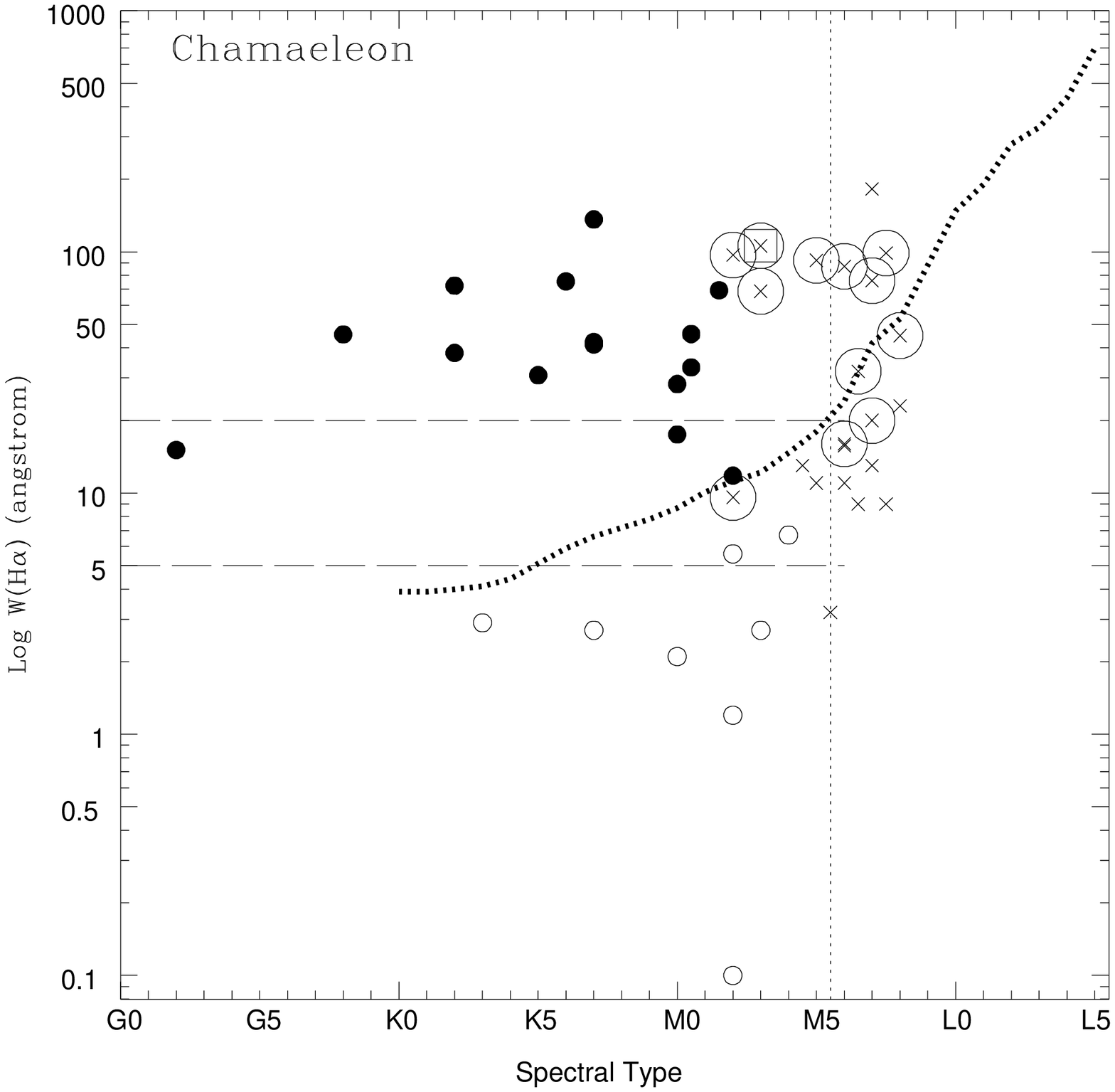} 
\caption{ (continue)} 
         \label{} 
   \end{figure*} 
%  ______________________________________________________________       
 
\clearpage 
%  %  %  %  %  %  %  %  %  %  %  %  %  %  %  %  %  %  %  %  %  %  %  %  %  % 
%  %  %  %  %  %  %  %  %  %  %  %  %  %  %  %  %  %  %  %  %  %  %  %  %  % 
%  %  %  %  %  %  %  %  %  %  %  %  %  %  %  %  %  %  %  %  %  %  %  %  %  % 
 
\setcounter{figure}{3} 
%  ----------------------------------------------------------- 
   \begin{figure*} 
   \centering 
   \includegraphics[width=14.0cm]{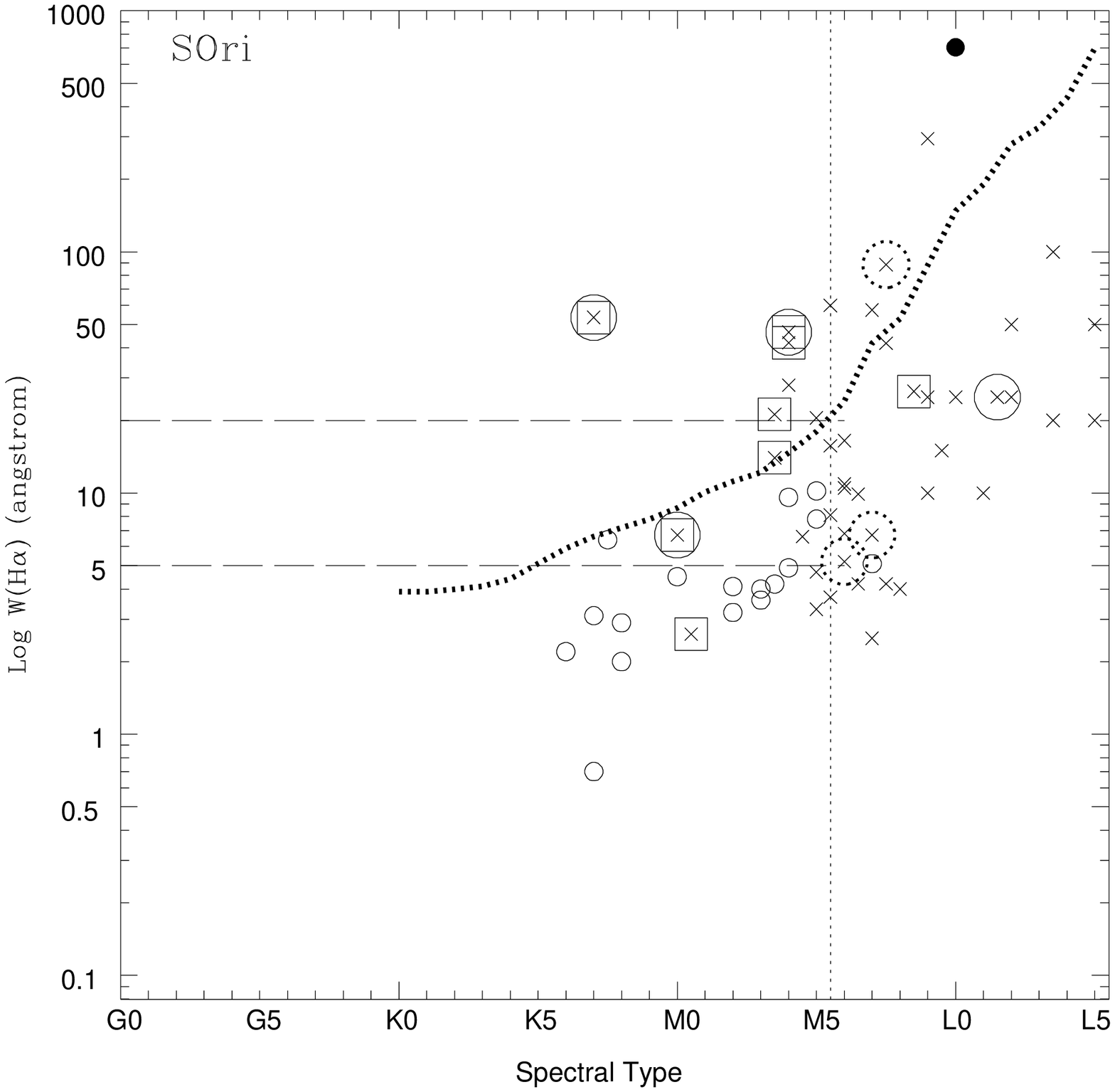} 
\caption{ (continue)} 
         \label{} 
   \end{figure*} 
%  ______________________________________________________________       
 
\clearpage 
%  %  %  %  %  %  %  %  %  %  %  %  %  %  %  %  %  %  %  %  %  %  %  %  %  % 
%  %  %  %  %  %  %  %  %  %  %  %  %  %  %  %  %  %  %  %  %  %  %  %  %  % 
%  %  %  %  %  %  %  %  %  %  %  %  %  %  %  %  %  %  %  %  %  %  %  %  %  % 
\setcounter{figure}{3} 
%  ----------------------------------------------------------- 
   \begin{figure*} 
   \centering 
   \includegraphics[width=14.0cm]{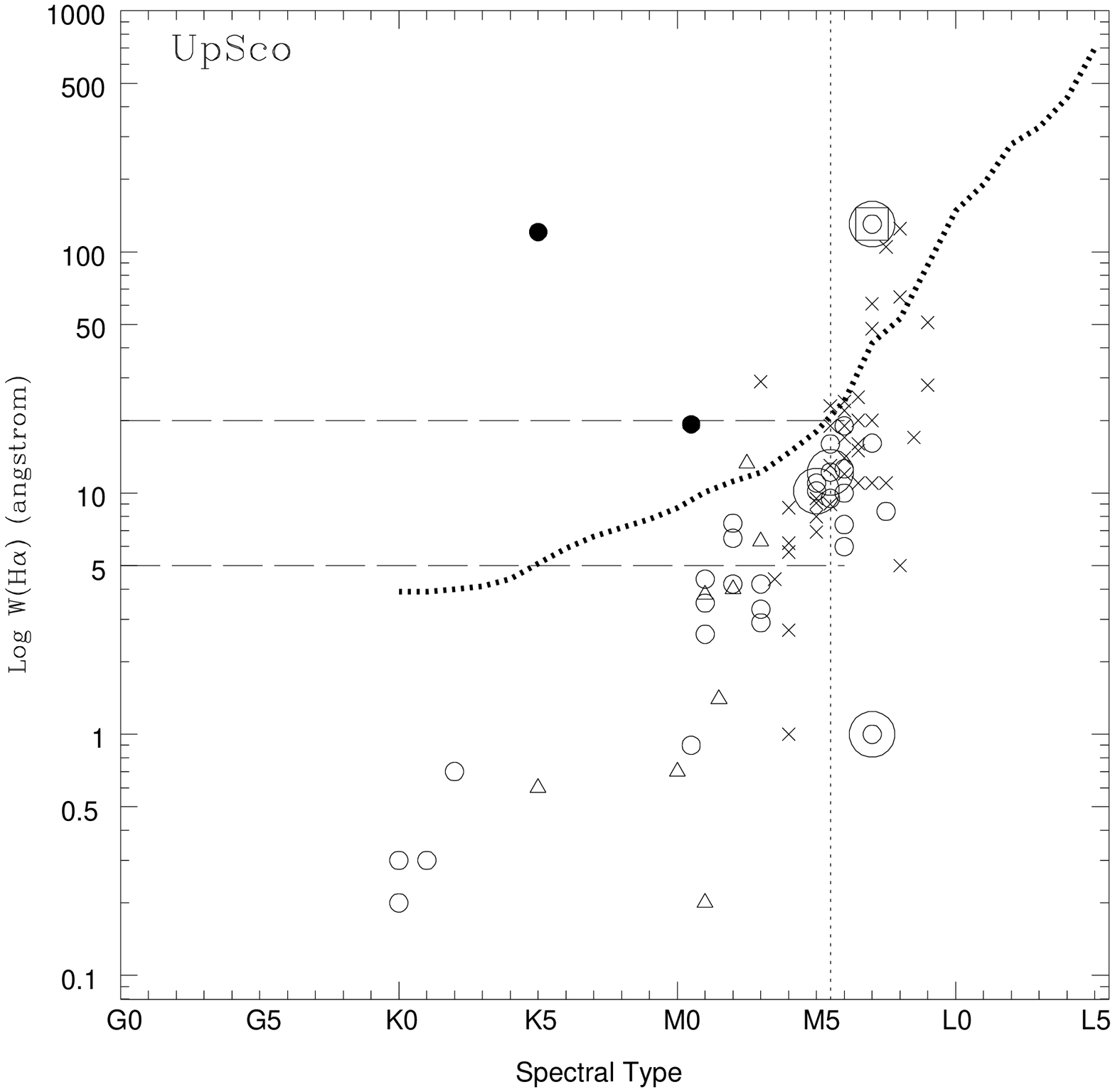} 
\caption{ (continue) 
} 
         \label{} 
   \end{figure*} 
%  ______________________________________________________________       
 
\clearpage 
%  %  %  %  %  %  %  %  %  %  %  %  %  %  %  %  %  %  %  %  %  %  %  %  %  % 
%  %  %  %  %  %  %  %  %  %  %  %  %  %  %  %  %  %  %  %  %  %  %  %  %  % 
%  %  %  %  %  %  %  %  %  %  %  %  %  %  %  %  %  %  %  %  %  %  %  %  %  % 
 
\setcounter{figure}{3} 
%  ----------------------------------------------------------- 
   \begin{figure*} 
   \centering 
   \includegraphics[width=14.0cm]{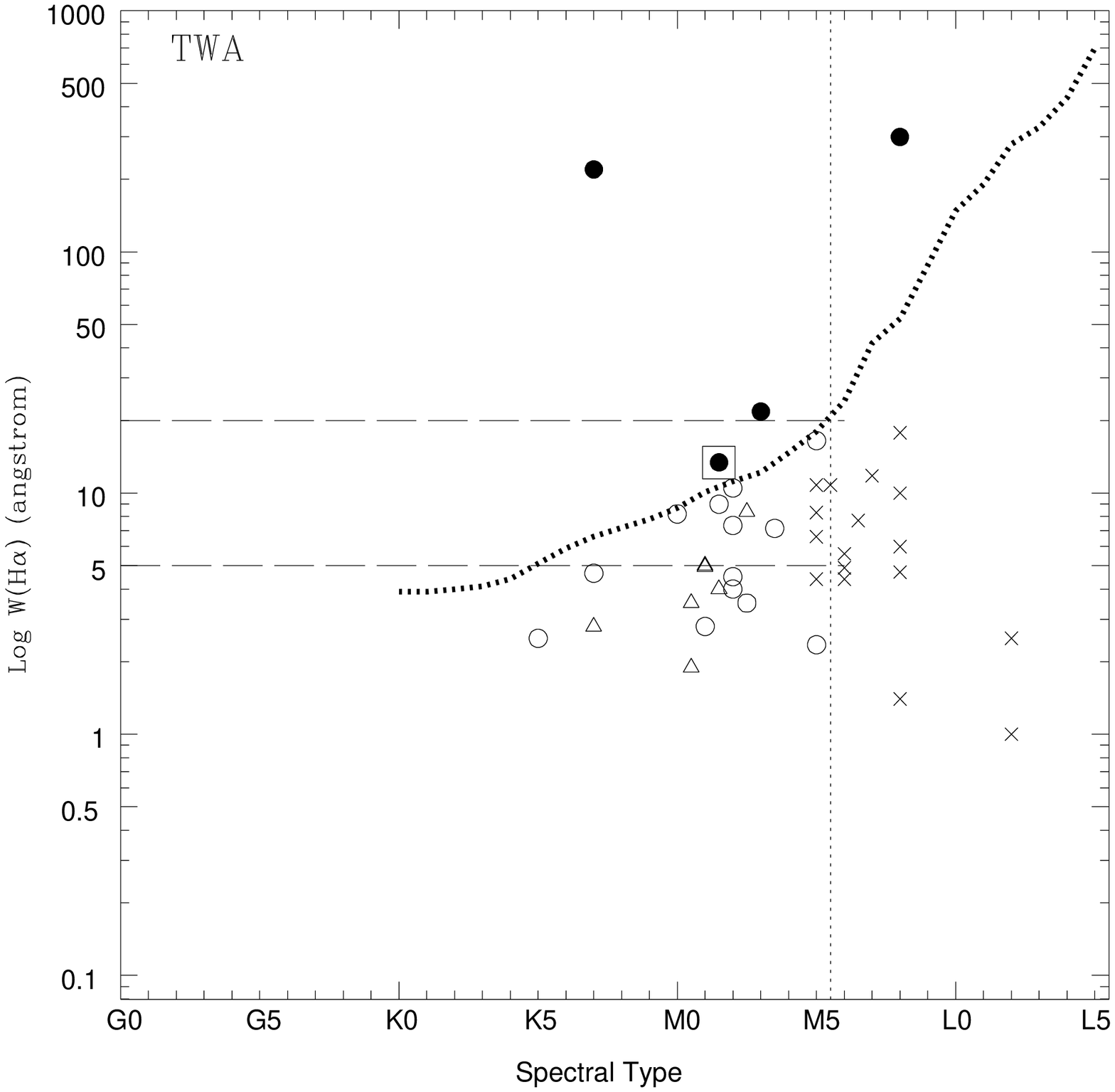} 
\caption{ (continue)} 
         \label{} 
   \end{figure*} 
%  ______________________________________________________________       
 
\clearpage 
%  %  %  %  %  %  %  %  %  %  %  %  %  %  %  %  %  %  %  %  %  %  %  %  %  % 
%  %  %  %  %  %  %  %  %  %  %  %  %  %  %  %  %  %  %  %  %  %  %  %  %  % 
%  %  %  %  %  %  %  %  %  %  %  %  %  %  %  %  %  %  %  %  %  %  %  %  %  % 

\setcounter{figure}{4} 
%  ----------------------------------------------------------- 
   \begin{figure*} 
   \centering 
   \includegraphics[width=14.0cm]{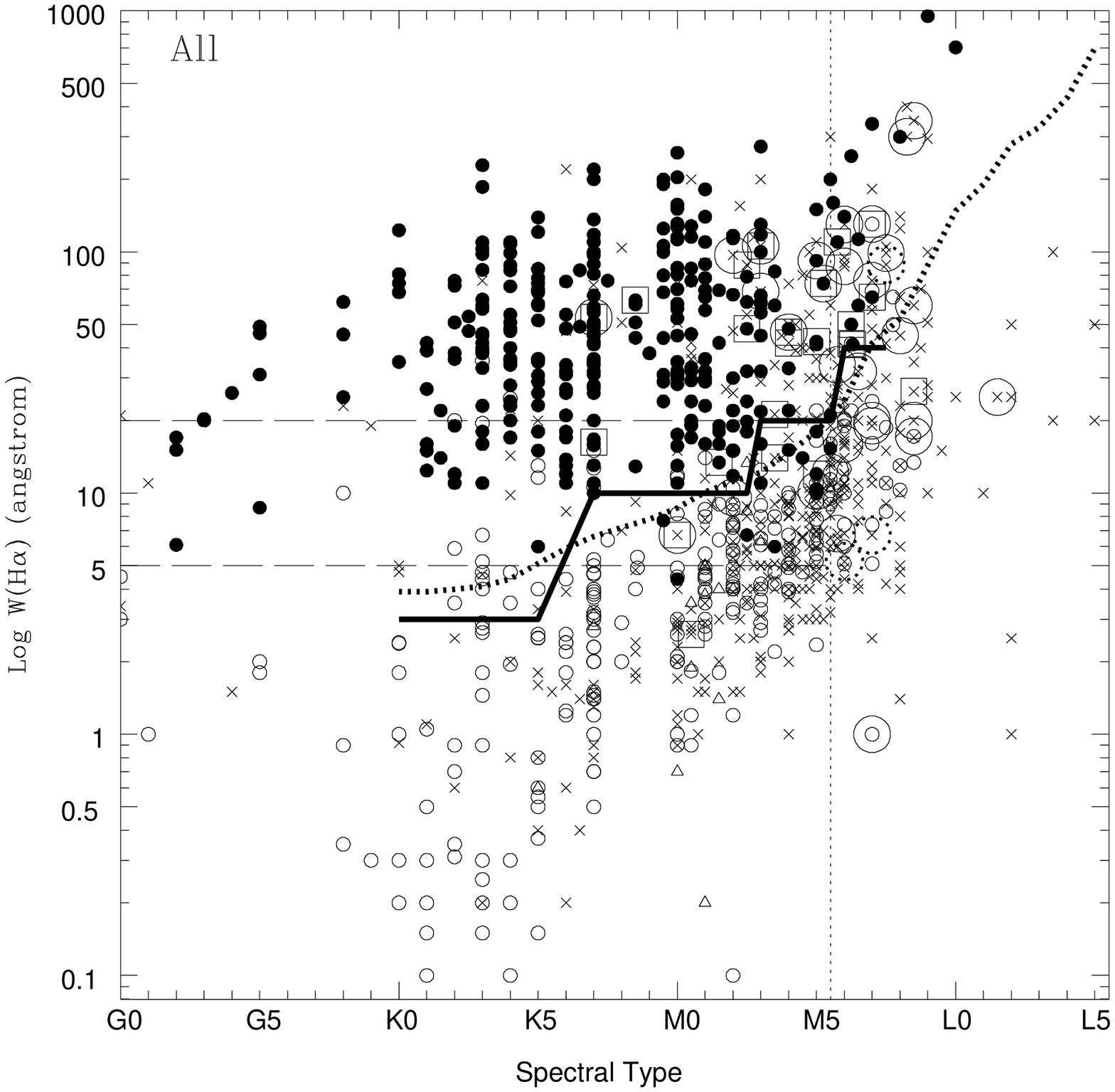} 
\caption{All data corresponding to the  
previous eight very young stellar associations. Labels as in Figure 4. 
We have also added the criterion defined by White \& Basri (2003) as a  
solid line.} 
         \label{} 
   \end{figure*} 
%  ______________________________________________________________       
 
\clearpage 
%  %  %  %  %  %  %  %  %  %  %  %  %  %  %  %  %  %  %  %  %  %  %  %  %  % 
%  %  %  %  %  %  %  %  %  %  %  %  %  %  %  %  %  %  %  %  %  %  %  %  %  % 
%  %  %  %  %  %  %  %  %  %  %  %  %  %  %  %  %  %  %  %  %  %  %  %  %  % 

\setcounter{figure}{5} 
%  ----------------------------------------------------------- 
\begin{figure*} 
\centering 
\includegraphics[width=14.0cm]{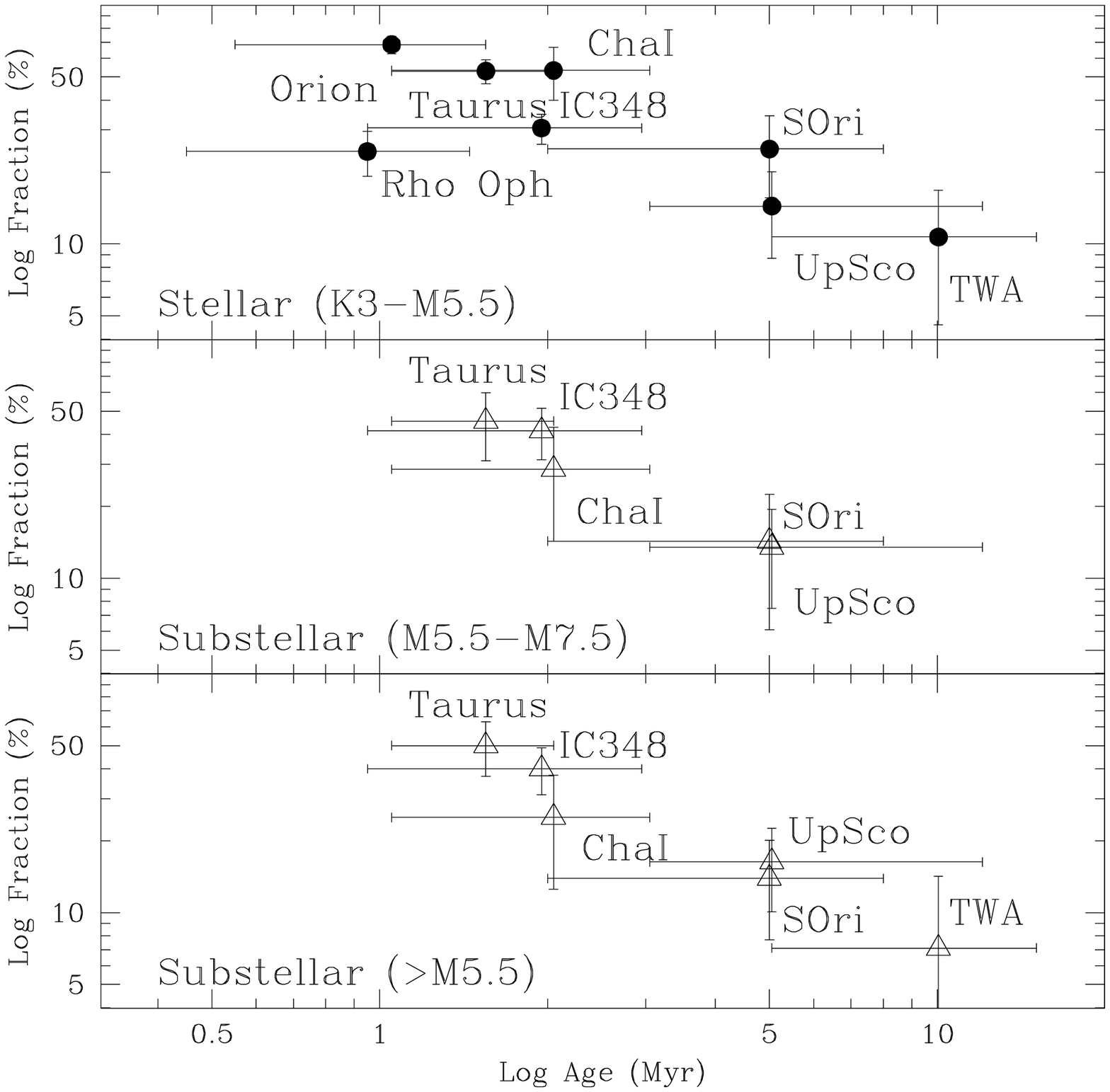} 
\caption{Fraction of objects classified as  
CTT stars or substellar analogs. Solid circles represent the 
stellar members, whereas the open triangles correspond to  
the substellar domain (M5.5--M7.5). Note the logarithmic scale. 
} 
         \label{} 
\end{figure*} 
%  ______________________________________________________________       

\clearpage 
%  %  %  %  %  %  %  %  %  %  %  %  %  %  %  %  %  %  %  %  %  %  %  %  %  % 
%  %  %  %  %  %  %  %  %  %  %  %  %  %  %  %  %  %  %  %  %  %  %  %  %  % 
%  %  %  %  %  %  %  %  %  %  %  %  %  %  %  %  %  %  %  %  %  %  %  %  %  % 

\setcounter{figure}{6} 
%  ----------------------------------------------------------- 
\begin{figure*} 
\centering 
\includegraphics[width=14.0cm]{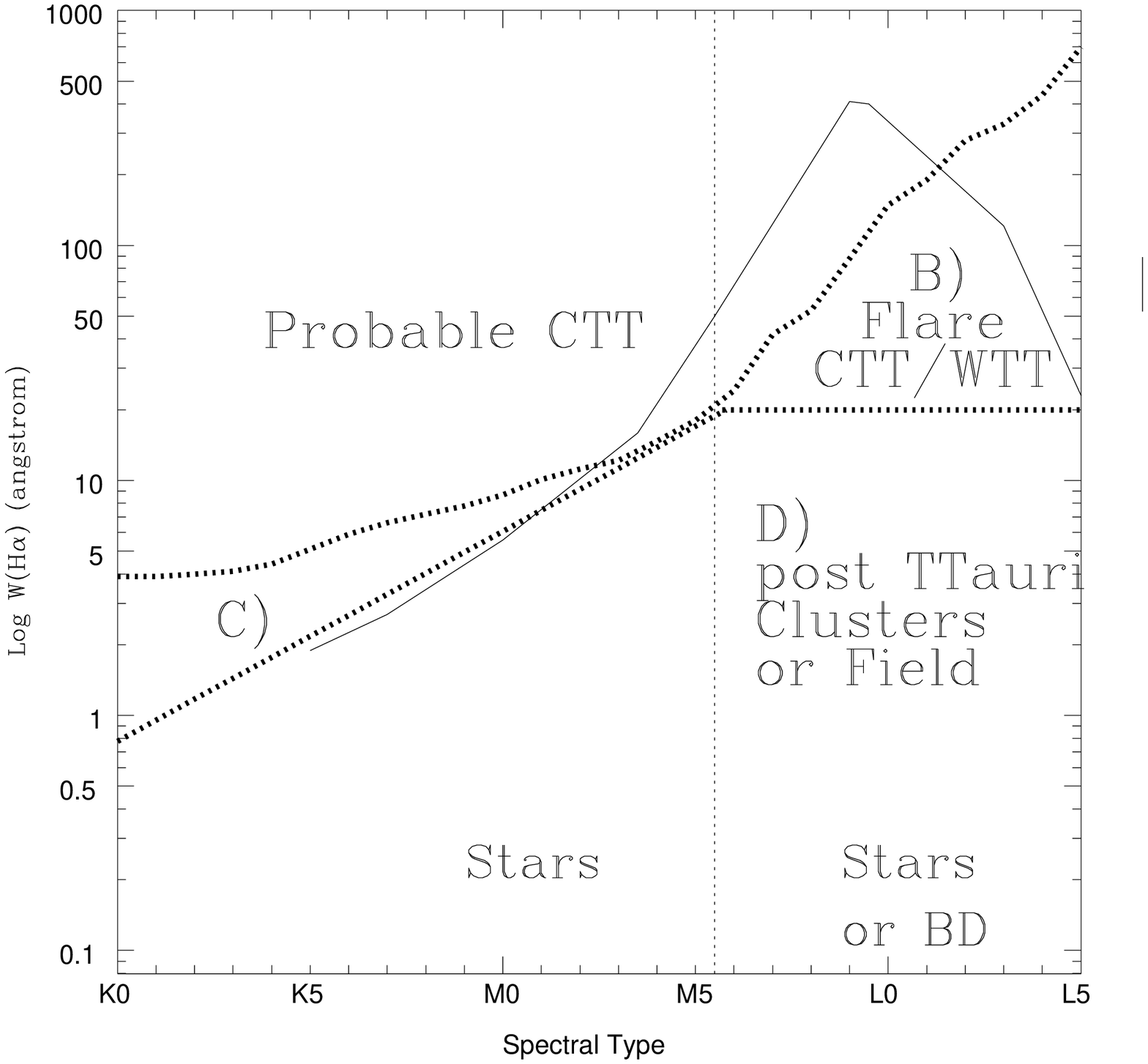} 
\caption{ Final criteria to classify CTTS and substellar analogs 
(thick, dotted lines, corresponding to the saturation limit and the maximum 
activity found in young open clusters). These lines divide the diagram in four very well 
defined areas. The thin solid line corrsponds to the maximum 
due to flares or very large variability. } 
         \label{} 
\end{figure*} 
%  ______________________________________________________________       

\clearpage 
%  %  %  %  %  %  %  %  %  %  %  %  %  %  %  %  %  %  %  %  %  %  %  %  %  % 
%  %  %  %  %  %  %  %  %  %  %  %  %  %  %  %  %  %  %  %  %  %  %  %  %  % 
%  %  %  %  %  %  %  %  %  %  %  %  %  %  %  %  %  %  %  %  %  %  %  %  %  % 

\end{document}